\documentclass[12pt,parskip=half, DIV=calc, BCOR=10mm, x11names]{scrbook}
\usepackage[l2tabu]{nag} 
\usepackage{amsmath}
\usepackage[usenames,dvipsnames]{xcolor}
\usepackage{empheq}
        
\usepackage{tikz}
\usepackage{pgfplots}
\usetikzlibrary{calc}
\usetikzlibrary{decorations.markings}
\usetikzlibrary{intersections}

\usetikzlibrary{fadings}
\tikzfading[name=fade right,
            left color=transparent!0,
            middle color=transparent!0,
            right color=transparent!100]
\tikzfading[name=fade left,
            left color=transparent!100,
            middle color=transparent!0,
            right color=transparent!0]
\tikzfading[name=fade up,
            top color=transparent!100,
            bottom color=transparent!0]
\tikzfading[name=fade down,
            top color=transparent!0,
            bottom color=transparent!100]

\newcommand\tempgraphs[1]{}

\usepackage[ilines, headsepline]{scrpage2}
\setheadwidth[0pt]{textwithmarginpar}

\clearscrheadfoot
\ihead{\headmark}
\ohead{\pagemark}

\addtokomafont{pagenumber}{\bfseries\Large\color{DeepSkyBlue4}}
\addtokomafont{pagehead}{\color{DeepSkyBlue4}}

\renewcommand{\textit}[1]{\textcolor{DeepSkyBlue4}{\emph{#1}}}

\usepackage{graphicx}
\usepackage[italian]{babel}
\usepackage{makeidx}
\usepackage{amsmath,amssymb,amsfonts}
\usepackage{enumitem}

\usepackage{lmodern}
\usepackage[T1]{fontenc}\usepackage{libertine}
\usepackage[utf8]{inputenc}
\usepackage[final, babel]{microtype} 

\usepackage{ifdraft}        
\usepackage{fixme}        

\usepackage{float}                      

\ifdraft{
\usepackage{draftwatermark} 
\SetWatermarkText{COPY~---~DRAFT}
\SetWatermarkAngle{90}
\SetWatermarkScale{6.0}
\SetWatermarkLightness{0.85}
\SetWatermarkFontSize{12 pt}
}{}

\numberwithin{equation}{section}

\usepackage{marginnote}
\newcommand{\mynote}[2][0pt]{}
\newcommand{\mynoteleft}[2][0pt]{}

\newcommand\answer\mynote

\renewcommand\thechapter{\Roman{chapter}}
\renewcommand\thesection{\arabic{section}}
\renewcommand\thesubsection{\arabic{section}.\arabic{subsection}}

\usepackage{titlesec}
\titleformat{\chapter}[display]
  {\normalfont\LARGE\sffamily}{\chaptertitlename\ \thechapter}{0pt}  
  {\titlerule\vskip2pt\titlerule\vskip20pt\Huge\bfseries\filleft}
\titleformat{\section}{\Large\sffamily\bfseries}{\thesection}{1em}{}
\titleformat{\subsection}{\large\sffamily\bfseries}{\thesubsection}{1em}{}

\usepackage{tabularx,graphicx,booktabs}

\usepackage{slantsc}

\usepackage{textcomp}

\newtheorem{Definizione}{Definizione}

\newtheorem{Teorema}{Teorema}
\newtheorem{Proprieta}{Propriet\`a}
\newtheorem{Esempio}{Esempio}

\newcounter{esercizio}
\renewcommand\theesercizio{\arabic{esercizio}} 
\newenvironment{Esercizio}{

\smallskip
\refstepcounter{esercizio}
\noindent\textsf{\bfseries Esercizio {\theesercizio}}\quad
\textsf\bgroup}
{\egroup\smallskip

}
\newlist{questions}{enumerate}{1}
\setlist[questions,1]{label=(\alph*)}

\newcommand{\eq}{\begin{equation}}
\newcommand{\beq}{\begin{equation}}
\newcommand{\eeq}{\end{equation}}
\newcommand{\bal}{\begin{align}}
\newcommand{\eal}{\end{align}}
\newcommand\bseq{\begin{subequations}}
\newcommand\eseq{\end{subequations}}
\newcommand{\definizione}[1]{{\begin{Definizione} {#1} \end{Definizione}}}
\newcommand{\esercizio}[1]{{\begin{Esercizio} #1 \end{Esercizio}}}

\newcommand{\newt}{G_\mathsf{N}}
\newcommand{\rhocrit}{\rho_{\mathsf{crit}}}
\newcommand{\rhocritzero}{\rho_{\mathsf{crit},0}}
\newcommand{\rhom}{\rho_{\mathsf{mat}}}
\newcommand{\rhor}{\rho_{\mathsf{rad}}}
\newcommand{\Mp}{M_{\mathsf{p}}}
\newcommand\press{P}
\newcommand\sphere{{\mc S}}

\newcommand{\lp}{\left(}
\newcommand{\rp}{\right)}
\newcommand{\mc}{\mathcal}

\newcommand{\calS}{{\mathcal S}}
\newcommand{\HH}{{\mathbb H}}
\newcommand{\RR}{{\mathbb R}}

\newcommand{\dd}{\mathrm{d}}
\newcommand{\p}{\partial}

\newcommand{\al}{\alpha }

\newcommand{\ga}{\gamma }
\newcommand{\Ga}{\Gamma }
\newcommand{\ep}{\epsilon }

\newcommand{\la}{\lambda }
\newcommand{\om}{\omega }

\newcommand{\La}{\Lambda }
\newcommand{\Si}{\Sigma }
\newcommand{\Om}{\Omega }
\newcommand{\munu}{{\mu\nu}}

\newcommand{\commentout}[1]{}

\makeindex

\begin{document}
\pagestyle{scrheadings}


\title{Introduzione alla cosmologia}
\author
{
Marco~M.~Caldarelli\footnote
{
\texttt{M.M.Caldarelli@soton.ac.uk}
}
\\
\normalsize
Mathematical Sciences and STAG Research Centre,\\
\normalsize
University of Southampton, United Kingdom
}
\date{Aprile 2016}
\maketitle

\tableofcontents


\chapter{Cosmologia}

\section{Il principio cosmologico e la geometria dell'universo}\label{sec::principio}

Nell'ambito della relativit\`a generale, lo spazio-tempo \`e una variet\`a lorentziana quadridimensionale, la cui geometria \`e determinata dalla distribuzione di materia in esso contenuta tramite le equazioni di Einstein
\begin{equation}
G_{ab}+\Lambda g_{ab}=8\pi\newt\, T_{ab}\,.
\label{c-einstein}
\end{equation}
Risulta naturale perci\`o chiedersi quale sia la soluzione di queste equazioni che descrive fedelmente l'universo nel quale viviamo. Questo, insieme alla descrizione dell'evoluzione dell'universo che l'ha portato ad essere come lo osserviamo oggi, \`e l'oggetto di studio della cosmologia.
Per risolvere le equazioni di Einstein dobbiamo conoscere il membro di destra, e quindi formulare ipotesi sulla distribuzione della materia nell'universo.
Questo non \`e un compito facile, in quanto abbiamo accesso solamente a osservazioni in una piccola parte del nostro cono luce passato.
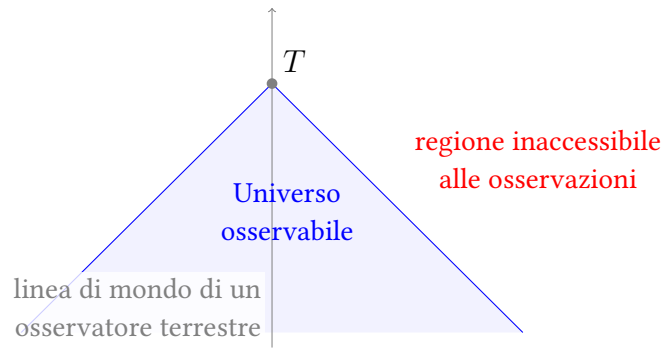
\begin{figure}
\centering
\begin{tikzpicture}
\fill[blue!5!white] (-3.3,-0.8)--(0,2.5)--(3.3,-0.8)--cycle;
\draw[blue] (-3.3,-0.8)--(0,2.5)--(3.3,-0.8);
\fill[white,opacity=0.8] (-3.3,-0.8)--(-3.3,0)--(-0.1,0)--(-0.1,-0.8)--cycle;
\draw[->, help lines] (0,-1.0) node [anchor=south east,align=right]
{
\small linea di mondo di un\\
\small osservatore terrestre
} -- (0,3.5);
\draw[red] (3.5,1.5) node[align=center]
{
\small regione inaccessibile\\
\small alle osservazioni
};
\draw[blue] (0.2,0.8) node[align=center]
{
\small Universo\\
\small osservabile
};
\fill[gray] (0,2.5) circle(2pt) node[black,anchor=south west]{$T$};
\end{tikzpicture}
\caption{\textbf{Cono luce passato.} Un osservatore sulla Terra ($T$) pu\'o osservare solamente eventi avvenuti nel suo cono luce passato, e dunque solo una regione finita dell'Universo \`e osservabile.}
\label{fig:conoLucePassato}\end{figure}
Qualunque ipotesi sul contenuto in materia dell'universo, che innalzeremo a rango di principio, comporter\`a un'ampia estrapolazione dei dati osservativi a cui abbiamo accesso. L'adeguatezza delle ipotesi e il successo dei modelli cosmologici che proporremo verranno misurati dalla capacit\`a del modello a spiegare lo stato presente dell'universo e dalla conferma osservativa delle sue predizioni.

\`E dalla rivoluzione copernicana che l'uomo ha abbandonato l'idea di occupare una posizione e un ruolo speciali nell'universo. Faremo nostro il \emph{principio copernicano}\index{principio!copernicano}, che afferma che non occupiamo una posizione geometricamente o fisicamente privilegiata nell'universo. In altre parole, in un qualunque altro punto dell'universo verificheremmo le stesse leggi della fisica e osserveremmo le stesse propriet\`a dell'universo. Questa enorme estrapolazione potr\`a essere giustificata solo dal successo del modello nello spiegare le osservazioni odierne.

A prima vista, il principio copernicano sembra paradossale in quanto, localmente, l'universo che osserviamo \`e disomogeneo: nel cielo notturno, quando siamo fortunati, vediamo la Via Lattea, e la materia luminosa appare concentrata nelle stelle. Queste si raggruppano in una gerarchia di strutture via via pi\'u estese -- galassie, ammassi di galassie, e super-ammassi separati tra loro da vuoti -- ma tutt'altro che omogenee (le scale di queste strutture sono indicate in tabella~\ref{tab:cosmicscales}).
\begin{table}[b]
\caption{\textbf{Scale delle strutture osservabili nel nostro universo.}}
\label{tab:cosmicscales}
\centering{\begin{tabularx}{.9\textwidth}{X c c}
\toprule
Oggetto & Massa ($M_\odot$) & Dimensioni ($\textsf{Mpc}$)\\
\midrule
Galassia a spirale & $10^{11}$ & $10^{-2}$\\
Distanza tra galassie vicine & --- & $1$\\
Ammassi di galassie & $10^{13}$ & $1$\\
Super-ammassi & $10^{15}$ & $10$\\
Vuoti & --- & $10$-$100$\\
Universo osservabile & $10^{22}$ & $10^{4}$\\
\bottomrule
\end{tabularx}}
\end{table}
Tuttavia, le osservazioni pi\'u recenti mostrano che su larga scala la distribuzione della materia nell'universo \`e \textit{omogenea}. L'evidenza di questo fatto viene dal conteggio delle galassie lontane, e dal fondo cosmico di raggi X e raggi $\gamma$. Inoltre, il cielo come viene osservato dalla Terra appare uguale in tutte le direzioni: \`e \textit{isotropo}. Anche qui, \`e solo sulle scale pi\'u estese che questa isotropia diventa manifesta, e la sua evidenza pi\'u forte e diretta proviene dalla scoperta della \emph{radiazione cosmica di fondo}\index{radiazione cosmica di fondo}\index{CMB|see{radiazione cosmica di fondo}} (Penzias e Wilson, 1964).
L'universo \`e permeato da una radiazione di corpo nero a $2.7\,\mathsf{K}$, estremamente isotropa. In effetti, le anisotropie che si misurano sono dell'ordine di una parte su diecimila.

\paragraph{Osservazioni cosmologiche.} Pi\'u in dettaglio, le osservazioni alla base della cosmologia moderna sono le seguenti:
\begin{enumerate}
\item \textsf{Distribuzione su larga scala della materia luminosa.} Su scale superiori a\footnote{Unit\`a di misura adeguate allo studio di oggetti astronomici e cosmologici sono l'\textit{unit\`a astronomica}\index{unit\`a astronomica}\index{$\textsf{ua}$|see{unit\`a astronomica}} ($\textsf{ua}$), che corrisponde alla distanza media Terra-Sole, $1\,\textsf{ua}=149\,600\,000\,\textsf{km}$, e il \textit{parsec}\index{parsec}\index{$\textsf{pc}$|see{parsec}} ($\textsf{pc}$), definito come la distanza dalla Terra (o dal Sole) di una stella che ha una parallasse annua di 1 secondo d'arco. In anni luce\index{anno luce}\index{$\textsf{al}$|see{anno luce}} (\textsf{al}), abbiamo $1\text{pc}\approx3.26\,\text{a.l.}$ In cosmologia, \`e conveniente misurare le distanze in \textsf{Mpc}, $1\textsf{Mpc}\approx3.09\cdot10^{24}\text{cm}$.}
$100\,\textsf{Mpc}$ l'universo appare liscio e non si osservano strutture di dimensioni superiori a quelle dei super-ammassi.
\item \textsf{Legge di Hubble.\footnote
{
Questa legge venne derivata dalla relativit\`a generale da George Lema\^\i tre nel 1927, e confermata due anni dopo dalle osservazioni di Edwin Hubble.
}
} Attraverso la misura dello spostamento verso il rosso (o \emph{redshift}\index{redshift}) dello spettro di emissione/assorbimento degli atomi che formano le stelle, si ottiene la velocit\`a relativa a noi delle altre galassie. Si osserva che si allontanano tutte dalla nostra galassia, con una velocit\`a di recessione\index{velocit\`a di recessione} $v=Hr$ proporzionale alla loro distanza $r$. Il fattore di proporzionalit\`a \`e la costante di Hubble $H_0$\index{Hubble!costante di}\index{$H_0$|see{Hubble, costante di}}, il cui valore attuale\footnote{In cosmologia, indichiamo con uno $0$ in pedice il valore attuale di una data quantit\`a fisica.} \`e 
$H_0=67.3\pm1.2\,\textsf{km}/(\textsf{s}\cdot\textsf{Mpc})$.
L'universo \`e dunque dinamico, in continua espansione!

Possiamo ottenere una stima dell'et\`a dell'universo\index{universo!et\`a dell'} osservando che $(H_0)^{-1}$ ha le dimensioni di un tempo. Questa quantit\`a si dice \emph{tempo di Hubble}\index{Hubble!tempo di}, e una rapida conversione produce un valore di circa 14.5 miliardi di anni. Questa \`e la prima indicazione che nel passato l'universo era molto piccolo e caldo, e si \`e poi raffreddato  durante la sua espansione (analogamente a quanto accade a un gas in espansione adiabatica). Questo spiega la terza osservazione:
\item \textsf{Radiazione cosmica di fondo.} Il nostro universo \`e permeato da una radiazione di corpo nero (con spettro di Planck) a temperatura $T=2.7255(6)\textsf{K}$, scoperta da Penzias e Wilson nel 1964. Questa radiazione \`e prevista dall'espansione cosmica: l'universo inizialmente era molto caldo e denso, e le elevate energie in gioco -- superiori all'energia di legame dell'atomo d'idrogeno -- mantenevano la materia nello stato di plasma all'equilibrio termico. I fotoni interagivano fortemente con questo plasma formato da particelle cariche (protoni ed elettroni). Il cammino libero medio di questi fotoni era piccolo, e l'universo era dunque opaco alla radiazione elettromagnetica.
All'\textit{epoca della ricombinazione\index{epoca della ricombinazione}}, quando l'et\`a dell'universo era di circa 372\,000 anni, la temperatura del plasma \`e scesa a $T\approx3\,000K$, e i protoni riescono finalmente a catturare elettroni in modo stabile, per formare i primi atomi di idrogeno. A quel punto, l'universo diventa trasparente ai fotoni, che da allora hanno viaggiato indisturbati fino ad oggi. Questi fotoni che osserviamo formano la radiazione cosmica di fondo\index{radiazione cosmica di fondo} (CMB), e provengono dalla radiazione di corpo nero presente al momento del disaccoppiamento. Abbiamo dunque accesso a una vera e propria fotografia dell'universo come era a quell'epoca! Mappe dettagliate della CMB sono state ottenute con strumenti in orbita (COBE, WMAP, Planck) e ne confermano l'omogeneit\`a e l'isotropia con un alta precisione: le anisotropie relative misurate sono $\Delta T/T\approx10^{-5}$ (vedi figura~\ref{fig:cmb}). Dallo spettro delle anisotropie si possono estrarre numerose informazioni sul nostro universo. Ad esempio, \`e possibile verificare che le sezioni dell'universo sono piatte, e misurare l'abbondanza relativa di barioni ed energia oscura.
\begin{figure}
\centerline
{
\includegraphics[width=0.8\textwidth]{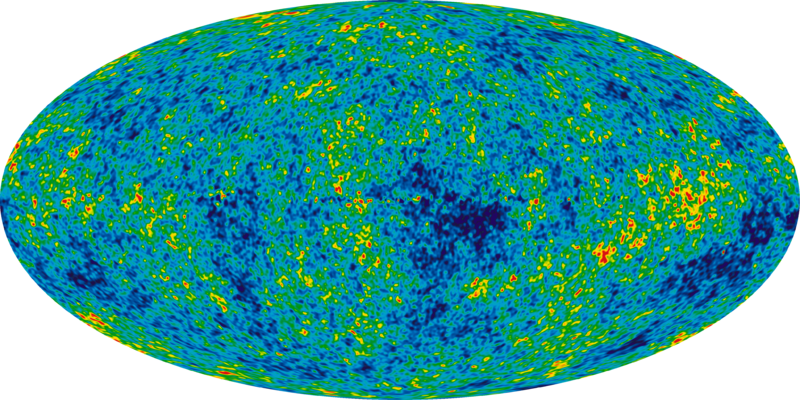}
}
\caption
{\textbf{
Mappa della radiazione cosmica di fondo.} Questa immagine, proveniente dai dati raccolti in nove anni da \textsf{WMAP} (2012), mostra le fluttuazioni di temperatura ($\pm200\mathsf{\mu K}$) al momento del disaccoppiameto, 13.77 miliardi di anni fa.
Queste fluttuazioni corrispondono alle disomogeneit\`a della densit\`a dalle quali sono cresciute le galassie.
\textsf{\footnotesize
Credit:
NASA/WMAP Science Team
}
}
\label{fig:cmb}
\end{figure}
\item \textsf{Osservarzione delle supernove di tipo Ia.}
Questi eventi molto luminosi permettono di misurare con precisione la relazione tra redshift e distanza di galassie molto lontane. Risulta una deviazione dalla legge Hubble; l'espansione dell'universo \`e \textit{accelerata}\index{espansione!accelerata}. Questa accelerazione \`e impressa dalla presenza di \textit{energia oscura}\index{energia oscura} (o \emph{dark energy}), la cui natura \`e tutt'ora un importante problema aperto, bench\'e corrisponda al 68.5\% 
dell'energia presente nell'universo. Le propriet\`a fisiche di questa energia oscura sono compatibili con quelle di una \textit{costante cosmologica positiva}\index{costante!cosmologica}, $\Lambda>0$.
\item\textsf{Composizione della densit\`a di energia dell'universo.} L'energia oscura forma la parte predominante del contenuto del nostro universo. Il 68.5\% della densit\`a di energia attualmente presente consiste in effetti di energia oscura. Un ulteriore 26.5\% della densit\`a di energia totale \`e dovuta alla materia oscura,
\index{materia oscura}
di cui si conoscono alcune propriet\`a (ad esempio che \`e composta da particelle non barioniche, non relativistiche, e che interagiscono soltanto attraverso la forze gravitazionali e deboli), la cui natura non \`e ancora nota. Solamente il rimanente 5\% \`e dovuto alla materia ordinaria. Questa, composta da barioni, \`e l'unica componente luminosa che gli astronomi possono osservare direttamente. In altre parole, abbiamo accesso solo una piccola frazione della materia contenuta nell'universo, che \`e composto al 95\% di energia e materia di cui non conosciamo l'origine! Vi sono infine piccole componenti di radiazione (la CMB) e di materia ultra-relativistica (neutrini) assimilabile a radiazione. Torneremo su questi contributi pi\'u avanti. 
\item\textsf{L'epoca inflazionaria.}
Molte propriet\`a dello spettro delle fluttuazioni della radiazione cosmica di fondo vengono spiegate supponendo che l'universo abbia subito, durante i suoi primissimi istanti di vita, una fase di espansione esponenziale chiamata \textit{inflazione}\index{inflazione}. Questo processo, che studieremo in dettaglio pi\'u avanti, \`e all'origine dell'estrema omogeneit\`a e piattezza spaziali che osserviamo. Le misure della polarizzazione della CMB forniscono gi\`a dei vincoli sul meccanismo dettagliato dell'inflazione, e in futuro misure pi\'u precise potrebbero portare a una prova incontrovertibile che l'epoca inflazionaria \`e effettivamente avvenuta.
\end{enumerate}

\paragraph{Il principio cosmologico.}
Le ipotesi che faremo nello studio della cosmologia sono dunque:
(a) non occupiamo una posizione privilegiata nell'universo (principio copernicano),
(b) l'universo che osserviamo dalla Terra \`e isotropo.
Il principio copernicano ci dice che in ogni istante $t$, ogni punto dello spazio deve essere equivalente a ogni altro punto, e dunque che l'universo \`e \textbf{spazialmente} omogeneo (ma le osservabili fisiche possono cambiare nel tempo: l'universo \`e in espansione e dunque lo spazio-tempo stesso non \`e omogeneo).
Inoltre, per il principio copernicano, l'isotropia che osserviamo non \`e dovuta a una nostra posizione privilegiata, e dunque le intere sezioni spaziali a tempo costante sono isotrope. Segue dunque il \textit{principio cosmologico}\index{principio!cosmologico}: \textbf{l'universo \`e spazialmente omogeneo e isotropo}. Definire ora questi concetti in modo pi\'u rigoroso, in modo da poterli usare per vincolare la geometria dell'universo in cui viviamo.
\definizione{
Uno spazio-tempo si dice spazialmente omogeneo\index{spazio-tempo!spazialmente omogeneo|see{omogeneit\`a spaziale}}\index{omogeneit\`a spaziale} se lo posso foliare in una famiglia a un parametro di ipersuperfici di tipo spazio $\Sigma_t$ tali che, per ogni $t$ e fissati due punti $p,q\in\Sigma_t$, esiste un'isometria di $\Sigma_t$ che porta $p$ in $q$.
}
L'isotropia \`e pi\'u delicata, in quanto \`e un concetto che dipende dall'osservatore. In effetti, essa \`e manifesta in un determinato evento al pi\'u a un osservatore. Un secondo osservatore, in moto relativo rispetto al primo, misurerebbe un'anisotropia dallo stesso evento spazio-temporale (si pensi ad esempio a una distribuzione omogenea di materia infinitamente estesa in $\RR^3$; solo un osservatore a riposo rispetto a questa materia osserverebbe una distribuzione isotropa, un osservatore in moto relativo vedrebbe un'anisotropia dovuta al flusso della materia in una determinata direzione).
\definizione{Un'osservatore che si muove lungo una curva di tipo tempo $\mc C$ con quadrivelocit\`a $u^\mu$ si dice \textit{osservatore isotropo}\index{osservatore isotropo} se in ogni evento $p\in\mc C$ e per ogni coppia di vettori $v$, $w\in T_p\mc M$ ortogonali alla velocit\`a ($u\cdot v=u\cdot w=0$), esiste un'isometria dello spazio-tempo che lascia fissi $p$ e $u$, e che ruota $v$ in $w$.}
Se vi \`e un osservatore isotropo che passa per ogni punto dello spazio-tempo, allora lo spazio-tempo si dice spazialmente isotropo.
\definizione{Uno spazio-tempo $\mc M$ si dice \textit{spazialmente isotropo}\index{spazio-tempo!spazialmente isotropo} se esiste una congruenza di osservatori isotropi che copre tutto lo spazio-tempo.}
Mostreremo a breve che in un universo spazialmente omogeneo e isotropo le superfici di omogeneit\`a $\Sigma_t$ sono ortogonali al campo di velocit\`a $u^\mu$ degli osservatori isotropi, altrimenti sarebbe possibile costruire un vettore spaziale privilegiato proiettando $u^\mu$ sulle superfici di omogeneit\`a.
Per l'ipotesi di omogeneit\`a spaziale, l'esistenza di un singolo osservatore isotropo implica che l'universo \`e anche spazialmente isotropo. Viceversa, si pu\`o mostrare che se vi \`e un osservatore isotropo in ogni punto, allora l'universo \`e spazialmente omogeneo.
Infine, una propriet\`a importante degli spazi isotropi \`e che in essi \`e non vi \`e nessuna direzione geometricamente preferita, e dunque \`e impossibile costruire un vettore privilegiato ortogonale a $u^\mu$: Tutte le osservabili vettoriali e tensoriali devono potersi scrivere esclusivamente in termini della velocit\`a $u^\mu$ e della metrica $g_{\mu\nu}$. Viviamo in un universo estremamente simmetrico, e quest'ultima propriet\`a vincola fortemente la forma della sua metrica.

\paragraph{Scelta delle coordinate cosmiche.}
\begin{figure} 
   \centering
   \begin{tikzpicture}
\fill[red!3!white, opacity=.8] (0,0)
 .. controls (2,.2) and (4.5,.2) .. (6,0)
 .. controls (6.3,.5) and (6.4,1) .. (7,1.5)
  .. controls (6,1.7) and (3,1.8) .. (1,1.5)
   .. controls (.6,1) and (.4,.8) .. (0,0);
\draw[name path=t1, purple] (0,0) node[black,thin,anchor=south west]{\small $\;\;\Sigma_{\tau_1}$}
 .. controls (2,.2) and (4.5,.2) .. (6,0)
 .. controls (6.3,.5) and (6.4,1) .. (7,1.5)
  .. controls (6,1.7) and (3,1.8) .. (1,1.5)
   .. controls (.6,1) and (.4,.8) .. (0,0);
\fill[red!3!white, opacity=.8, shift={(.1,1.3)}] (0,0) 
 .. controls (2,.2) and (4.5,.2) .. (6,0)
  .. controls (6.3,.5) and (6.4,1) .. (7,1.5)
  .. controls (6,1.7) and (3,1.8) .. (1,1.5)
   .. controls (.6,1) and (.4,.8) .. (0,0);
\draw[name path=t2, purple,shift={(.1,1.3)}] (0,0) node[black,thin,anchor=south west]{\small $\;\;\Sigma_{\tau_2}$}
 .. controls (2,.2) and (4.5,.2) .. (6,0)
 .. controls (6.3,.5) and (6.4,1) .. (7,1.5)
  .. controls (6,1.7) and (3,1.8) .. (1,1.5)
   .. controls (.6,1) and (.4,.8) .. (0,0);

\draw (2.1,-1) node[anchor=west] {\small $\mc O_1$};
\draw (2.3,4) node[anchor=east] {\small $\tau$};
\draw[name path=pathA, help lines] (2.1,-1) .. controls (2,1) and (2.2,3) .. (2.3,4);
\begin{scope}
\clip[name intersections={of=pathA and t1}] (0,-1) rectangle ($(intersection-1)+(1,0)$);
\draw[black, thick] (2.1,-1) .. controls (2,1) and (2.2,3) .. (2.3,4);
\end{scope}
\begin{scope}
\clip[name intersections={of=pathA and t2}] (0,.5) rectangle ($(intersection-1)+(1,0)$);
\draw[black, thick] (2.1,-1) .. controls (2,1) and (2.2,3) .. (2.3,4);
\end{scope}
\begin{scope}
\clip(0,2) rectangle (3,4.5);
\draw[black, thick,->] (2.1,-1) .. controls (2,1) and (2.2,3) .. (2.3,4);
\end{scope}
\path[name path=p1] (0,.5)--(5,.5);
\path[name path=p2] (0,2)--(5,2);
\draw[name intersections={of=pathA and p1}] (intersection-1) node[anchor=west] {\small $(\tau_1,x_1)$};
\draw[name intersections={of=pathA and p1}, help lines] ($(intersection-1)+(-2pt,-1pt)$)--($(intersection-1)+(2pt,1pt)$);
\draw[name intersections={of=pathA and p1}, help lines] ($(intersection-1)+(-2pt,1pt)$)--($(intersection-1)+(2pt,-1pt)$);
\draw[name intersections={of=pathA and p2}] (intersection-1) node[anchor=west] {\small $(\tau_2,x_1)$};
\draw[name intersections={of=pathA and p2}, help lines] ($(intersection-1)+(-2pt,-1pt)$)--($(intersection-1)+(2pt,1pt)$);
\draw[name intersections={of=pathA and p2}, help lines] ($(intersection-1)+(-2pt,1pt)$)--($(intersection-1)+(2pt,-1pt)$);

\draw (4.1,-1) node[anchor=west] {\small$\mc O_2$};
\draw[name path=pathB, help lines] (4.1,-1) .. controls (4,1) and (4.2,3) .. (4.3,4);
\begin{scope}
\clip[name intersections={of=pathB and t1}] (0,-1) rectangle ($(intersection-1)+(1,0)$);
\draw[black, thick] (4.1,-1) .. controls (4,1) and (4.2,3) .. (4.3,4);
\end{scope}
\begin{scope}
\clip[name intersections={of=pathB and t2}] (0,.8) rectangle ($(intersection-1)+(1,0)$);
\draw[black, thick] (4.1,-1) .. controls (4,1) and (4.2,3) .. (4.3,4);
\end{scope}
\begin{scope}
\clip(0,2.3) rectangle (6,4.5);
\draw[black, thick, ->] (4.1,-1) .. controls (4,1) and (4.2,3) .. (4.3,4);
\end{scope}
\path[name path=p1] (0,.8)--(5,.8);
\path[name path=p2] (0,2.3)--(5,2.3);
\draw[name intersections={of=pathB and p1}] (intersection-1) node[anchor=west] {\small $(\tau_1,x_2)$};
\draw[name intersections={of=pathB and p1}, help lines] ($(intersection-1)+(-2pt,-1pt)$)--($(intersection-1)+(2pt,1pt)$);
\draw[name intersections={of=pathB and p1}, help lines] ($(intersection-1)+(-2pt,1pt)$)--($(intersection-1)+(2pt,-1pt)$);
\draw[name intersections={of=pathB and p2}, help lines] ($(intersection-1)+(-2pt,-1pt)$)--($(intersection-1)+(2pt,1pt)$);
\draw[name intersections={of=pathB and p2}, help lines] ($(intersection-1)+(-2pt,1pt)$)--($(intersection-1)+(2pt,-1pt)$);
\draw[name intersections={of=pathB and p2}] (intersection-1) node[anchor=west] {\small $(\tau_2,x_2)$};
\end{tikzpicture}
   \caption{\textbf{Scelta delle coordinate cosmiche.}}
   \label{fig:cosmic}
\end{figure}
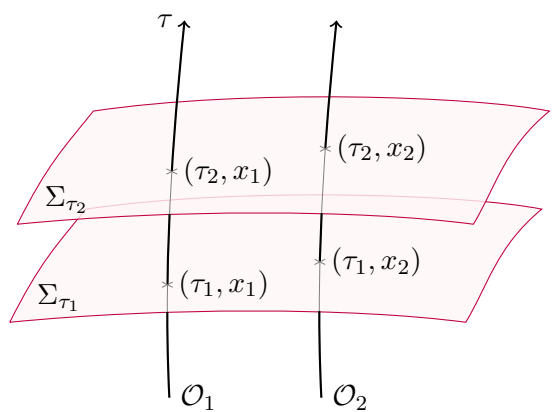
Per semplificare il problema, \`e conveniente scegliere coordinate in cui l'omogeneit\`a e l'isotropia spaziale diventano manifeste.
Prendiamo la linea di mondo di un osservatore isotropo, che chiameremo $\mc O_1$. Questa curva interseca ortogonalmente le ipersuperfici omogenee che foliano lo spazio-tempo. Usiamo il tempo proprio dell'osservatore $\mc O_1$ per definire il \textit{tempo cosmico}\index{tempo cosmico}, i.e.~chiamiamo $\Si_\tau$ l'ipersuperficie di omogeneit\`a intersecata da quest'osservatore quando il suo tempo proprio \`e $\tau$ (vedi figura~\ref{fig:cosmic}). Definiamo poi delle coordinate spaziali $\{x^i\}$ al tempo $\tau_1$, sulla ipersuperficie $\Si_{\tau_1}$. Gli osservatori isotropi formano una congruenza di curve che riempiono l'universo e dunque a
ogni punto $x_a\in\Si_{\tau_1}$ \`e associato un unico osservatore isotropo $\mc O_a$. Trasportiamo in questo modo la definizione delle coordinate spaziali a tutte le ipersuperfici di omogeneit\`a, in modo tale che le curve degli osservatori isotropi siano le curve $\{x^i=\text{costante}\}$.

Sia $u^\mu$ la quadri-velocit\`a degli osservatori isotropi ($u^\mu u_\mu=-1$). Abbiamo visto che le ipersuperfici $\Sigma_\tau$ sono ortogonali a $u^a$, altrimenti la sua proiezione sullo spazio tangente $\Sigma_\tau$ definirebbe un vettore privilegiato, violando cos\'\i\ l'isotropia ipotizzata.
Inoltre, in queste coordinate, le componenti miste $g_{\tau i}$ della metrica definiscono un vettore nello spazio tangente a $\Si_\tau$, e devono devono dunque annullarsi, sempre per rispettare l'isotropia. Perci\`o lo spazio-tempo \`e della forma $\mathbb R\times\Si_\tau$ e, nelle coordinate cosmiche, la sua metrica assume la forma
\begin{equation}
ds^2=-d\tau^2+h_{ij}(\tau,x^k)\,dx^idx^j,
\label{metrica uni}\end{equation}
dove $h_{ij}(\tau,x^k)$ \`e la metrica indotta sull'ipersuperficie spaziale $\Sigma_\tau$. Si vede facilmente che una particella a riposo in queste coordinate rimane a riposo, e $\tau$ misura il suo tempo proprio. Per questo motivo, le coordinate $\{x^i\}$ sono dette coordinate \textit{comoventi}\index{coordinate!comoventi}.
\begin{Esercizio}\label{ex:ossgeo}
Mostrare che le curve $\{x^i=\text{costante}\}$ risolvono l'equazione delle geodetiche per la metrica \eqref{metrica uni}, e che la coordinata $\tau$ ne misura il tempo proprio.
\end{Esercizio}
Possiamo dunque riformulare il principio cosmologico\index{principio!cosmologico} nel modo seguente: \emph{\`e possibile definire nello spazio-tempo una famiglia di sezioni spaziali omogenee e isotrope, i.e. tali che su ciascuna di esse l'universo ha le stesse propriet\`a fisiche in ogni punto e direzione}.

\paragraph{Spazi a curvatura costante.}
Studiamo ora la geometria delle sezioni spaziali dell'universo. Devono formare spazi riemanniani tridimensionali omogenei e isotropi, il che limita ulteriormente la forma della metrica. Innanzitutto, per l'isotropia, le osservazioni che effettuiamo da un determinato punto $p$ dello spazio non dipendono dalla direzione di osservazione. Lo spazio \`e dunque invariante per rotazioni intorno a questo punto.
Per definire coordinate spaziali che rispettino questa simmetria sferica, adottiamo una strategia analoga a quella usata qui sopra. Prendiamo una geodetica $\mc C$ uscente da $p$, e per ogni punto di questa geodetica consideriamo la sua orbita sotto il gruppo $SO(3)$ delle rotazioni. Questa orbita forma una sfera $\mc S^2$, e usiamo la sua area per definire una coordinata radiale\footnote{Questa coordinata \`e definita in modo invariante, in quanto l'area della sfera \`e un invariante sotto diffeomorfismi, ed \`e analoga alla coordinata di Schwarzschild che introdurremo pi\'u avanti nello studio delle configurazioni a simmetria sferica.}
$r=(\mc A/4\pi)^{1/2}$.
Infine, definiamo su una di queste sfere delle coordinate sferiche $(\theta,\phi)$, che trasportiamo su tutto lo spazio richiedendo che le geodetiche uscenti da $p$ siano curve con $\theta$ e $\phi$ costanti.
In queste coordinate, la metrica dello spazio isotropo deve essere della forma
\begin{equation}
d\sigma^2=f(r)\,dr^2+r^2\left(d\theta^2+\sin^2\theta\,d\varphi^2\right).
\label{c-metrica3d}\end{equation}
L'assenza di termini misti $g_{r\theta}$ e $g_{r\phi}$ proviene dall'ortogonalit\`a delle geodetiche radiali alle sfere: se cos\`i non fosse la proiezione del vettore tangente alle geodetiche sulla sfera definirebbe un vettore che non rimane invariante sotto rotazioni, in contraddizione con l'ipotesi di isotropia.

Rimane da determinare la forma della funzione $f(r)$.
La non esistenza di vettori privilegiati implica che il tensore di Riemann ${}^{(3)}R_{ijkl}$ di questi spazi deve potersi scrivere solamente in termini della loro metrica $h_{ij}$, unico oggetto tensoriale che abbiamo a disposizione. L'unica combinazione che condivide tutte le simmetrie del tensore di Riemann \`e la seguente:
\begin{equation}
{}^{(3)}R_{ijkl}=K(r,\theta,\phi)\left(h_{ik}h_{jl}-h_{il}h_{jk}\right),
\end{equation}
dove la funzione $K(r,\theta,\phi)$ \`e proporzionale alla curvatura scalare. Ma la curvatura scalare di uno spazio omogeneo \`e costante (altrimenti misurando la curvatura scalare potrei distinguere due punti $p$ e $q$ dello spazio) e dunque anche $K(r,\theta,\phi)=K$ \`e costante.
\definizione{Una variet\`a $\Sigma$ per cui $R_{ijkl}=2Kh_{i[k}h_{l]j}$ si dice \textit{spazio a curvatura costante}\index{spazio!a curvatura costante}.}
Contraendo il tensore di Riemann otteniamo il tensore di Ricci,
\begin{equation}
{}^{(3)}R_{ik}=2Kh_{ik}.
\label{c-ricci}\end{equation}
\esercizio{
Calcolare il tensore di Ricci per la metrica~\eqref{c-metrica3d} e mostrare che la soluzione dell'equazione~\eqref{c-ricci} \`e data da
\begin{equation}
f(r)=\frac1{1-Kr^2}\,.
\end{equation}
}
Gli spazi tridimensionali a curvatura costante hanno dunque una metrica della forma 
\begin{equation}
d\sigma^2=\frac{dr^2}{1-Kr^2}+r^2\left(d\theta^2+\sin^2\theta\,d\varphi^2\right),
\label{c-metrica3db}\end{equation}
con $K\in\RR$.
Se $K=0$, questo \`e lo spazio euclideo tridimensionale $\RR^3$, dotato della metrica piatta. Supponiamo ora $K\neq0$. Possiamo definire una nuova coordinata
\beq
\rho=\frac r{\sqrt{|K|}}
\eeq
che porta la metrica nella forma
\begin{equation}
d\sigma^2=\frac1{|K|}\left[\frac{dr^2}{1-kr^2}+r^2\left(d\theta^2+\sin^2\theta\,d\varphi^2\right)\right]\,,
\label{c-metrica3dc}\end{equation}
dove abbiamo definito $k=K/|K|=\pm1$. Se $k=+1$, la metrica \`e quella della 3-sfera\index{3-sfera} $\calS^3$, uno spazio a curvatura costante positiva, mentre se $k=-1$ la metrica descrive lo spazio iperbolico tridimensionale $\HH^3$\index{spazio!iperbolico}, spazio a curvatura costante negativa.
\begin{Esercizio}\label{es:quadriche}
\textbf{$\calS^3$ e $\HH^3$ come quadriche\index{quadriche}}\quad
Si consideri l'insieme dei punti $S=\{x_1^2+x_2^2+x_3^2+x_4^2=1\}$ in $\RR^4$ dotato di metrica euclidea $ds^2=dx_1^2+dx_2^2+dx_3^2+dx_4^2$. Definire delle coordinate su $S$ e trovarne la metrica indotta. Risolvere lo stesso problema per l'insieme di punti $H=\{x_1^2+x_2^2+x_3^2-x_4^2=1\}$. Verificare che $S$ e $H$ sono spazi a curvatura costante e calcolarne la curvatura.
\emph{Suggerimento: le relazioni $\cos^2\al+\sin^2\al=1$ e $\cosh^2\al-\sinh^2\al=1$ possono essere usate per definire le coordinate sulle quadriche.}
\mynote{Metrica indotta su sezioni coniche. Completare testo dell'esercizio.}
\end{Esercizio}
Dunque, a meno di una scala globale, vi sono solo tre scelte possibili per la 
geometria delle sezioni spaziali dell'universo. Tuttavia, questa scala che determina le dimensioni delle sezioni spaziali pu\`o dipendere dal tempo $\tau$ senza violare il principio cosmologico, $K=K(\tau)$. In definitiva, con una semplice ridefinizione, otteniamo la metrica,
\begin{equation}
ds^2=-d\tau^2+a^2(\tau)\left[
\frac{dr^2}{1-kr^2}+r^2\left(d\theta^2+\sin^2\theta\,d\varphi^2\right)\right]\,.
\end{equation}
Questa \`e la metrica di Robertson-Walker\index{metrica!di Robertson-Walker}, che descrive la geometria dell'universo.
Tutta la dinamica cosmologica viene ridotta in un'unica funziona $a(\tau)$, che si ricava dalle equazioni di Einstein una volta nota la distribuzione di materia nell'universo.
Questa funzione $a(\tau)$ si chiama \textit{fattore di scala}\index{fattore di scala}\index{$a(\tau)$|see{fattore di scala}}.

La costante $k$, che pu\`o solo assumere i tre valori discreti $0,$ $\pm1$, determina invece la natura della metrica spaziale:
\begin{itemize}
\item Per $k=+1$, le sezioni spaziali sono 3-sfere\index{3-sfera}. Questi sono spazi compatti a curvatura costante positiva e l'universo si dice \textit{chiuso}\index{universo!chiuso}.
\item Per $k=0$, le sezioni spaziali formano lo spazio euclideo $\RR^3$, ovvero spazi non compatti a curvatura nulla. L'universo si dice in questo caso \textit{piatto}\index{universo!piatto}.
\item Per $k=-1$, le sezioni spaziali sono spazi iperbolici\index{spazio!iperbolico} $\HH^3$, spazi non compatti a curvatura costante negativa. L'universo si dice \textit{aperto}\index{universo!aperto}.
\end{itemize}

Possiamo a questo punto porci le prime domanda di natura cosmologica: quale \`e la geometria locale e la geometria globale del nostro universo? ovvero se il nostro universo \`e compatto\footnote{Sia $\RR^3$ che $\HH^3$ possono essere resi compatti quozientandoli rispetto a un loro sottogruppo discreto di isometrie, analogamente a come si ottiene un toro (compatto) dal piano euclideo. Ci si pu\`o dunque chiedere se le sezioni spaziali sono semplicemente connesse. Tuttavia, una topologia non banale sembra esclusa dalle osservazioni, che impongono un limite inferiore di $10^{10}$ anni-luce alla lunghezza dei cicli (WMAP).} o infinitamente esteso, e se \`e aperto, piatto, o chiuso. \`E con le osservazioni che potremo rispondere a queste domande.

\`E utile definire una nuova coordinata radiale $\chi$ che misuri (a meno del fattore di scala $a(\tau)$) la lunghezza propria nelle sezioni spaziali,
\begin{equation}
r=S_k(\chi)=\left\{\begin{array}{ll}
\sin\chi  & \text{se $k=1$},\\
\chi      & \text{se $k=0$},\\
\sinh\chi & \text{se $k=-1$}.
\end{array}\right.
\label{CoordChi}\end{equation}
Nelle coordinate $(\tau,\chi,\theta,\varphi)$, la metrica di Robertson-Walker assume la forma
\begin{equation}
ds^2=-d\tau^2+a^2(\tau)\left[
d\chi^2+S^2_k(\chi)\left(d\theta^2+\sin^2\theta\,d\varphi^2\right)\right]\,.
\label{RWmetric}\end{equation}
Nel seguito di questo capitolo, useremo sempre questa forma della metrica.

Il fattore di scala $a(\tau)$ codifica la dinamica cosmologica, e la sua evoluzione si ottiene -- data la distribuzione di materia nell'universo --  risolvendo le equazioni di Einstein. Tuttavia, alcuni fenomeni sono puramente cinematici e si possono studiare indipendentemente dalla conoscenza della precisa dipendenza dal tempo del fattore di scala. Prima di risolvere le equazioni di Einstein e studiare la dinamica cosmologica, vediamo dunque cosa possiamo imparare dalla forma della metrica di Robertson-Walker.

\section{La cinematica cosmologica, il redshift cosmologico}
\paragraph{Moto degli osservatori isotropi.}
Le linee di mondo degli osservatori isotropi, $\{\chi,\theta,\varphi\}=\mathrm{cost.}$ sono geodetiche (vedi esercizio~\ref{ex:ossgeo}).
Per questo motivo, le coordinate di RW si dicono \emph{comoventi}, in quanto questi osservatori sono a riposo rispetto alla materia contenuta nell'universo (il \textit{fluido cosmologico}\index{fluido cosmologico}).
Dunque, le galassie si muovono lungo le linee di mondo degli osservatori isotropi, a meno di piccole fluttuazioni delle velocit\`a.

\paragraph{L'universo in espansione.}
Consideriamo due galassie, che possiamo supporre si trovino rispettivamente nei punti $(\chi_1,0,0)$ e $(\chi_2,0,0)$ senza alcuna perdita di generalit\`a. La loro distanza propria al tempo $\tau$ \`e data da $R=a(\tau)|\chi_2-\chi_1|$. Proprio per questo motivo abbiamo chiamato $a(\tau)$ `fattore di scala'. Se questo non \`e costante, segue che la distanza propria tra le due galassie varia con il tempo, alla velocit\`a
\beq
v=\frac{dR(\tau)}{d\tau}=\dot a(\tau)\left|\chi_2-\chi_1\right|=\frac{\dot a}{a}R(\tau).
\eeq
Qui, e per il resto di questo capitolo, con il punto indichiamo la derivata rispetto al tempo proprio $\tau$.
\`E naturale a questo punto definire il \emph{parametro di Hubble}\index{Hubble!parametro di}
\beq
H(\tau)=\frac{\dot a}a
\label{c-Hubble_parm}\eeq
dal quale segue una legge lineare per la velocit\`a di recessione (se $\dot a>0$, altrimenti si tratta di un avvicinamento; vedremo che l'osservazione del redshift delle galassie mostra che attualmente $\dot a$ \`e positivo)\index{Hubble!legge di},
\begin{equation}
v=HR.
\end{equation}
Questa variazione della distanza delle galassie mostra che il modello di Ro\-bert\-son-Walker non descrive un universo statico, bens\'\i\ un universo in \emph{espansione} o in \emph{contrazione}.
\`E bene fare a questo punto alcune osservazioni:
\begin{enumerate}
\item
Il parametro di Hubble $H(\tau)$ dipende dal tempo. Il suo valore attuale si indica con $H_0$ e si chiama \emph{costante di Hubble}\index{Hubble!costante di}. Bench\'e la sua misura sia soggetta a molte difficolt\`a sperimentali, abbiamo gi\`a visto che osservazioni recenti trovano
\beq
H_0=67.3\pm1.2\,\textsf{km}/(\textsf{s}\cdot\textsf{Mpc}).
\eeq
\item
Se $\dot a>0$ l'universo \`e in espansione, se invece
$\dot a<0$ l'universo \`e in una fase di contrazione.
\item
\`E bene comprendere correttamente la natura di questa espansione dell'universo.
\`E la scala delle distanze tra tutti gli osservatori isotropi (o galassie) che varia nel tempo: \`e lo spazio-tempo stesso che si espande o si contrae, non sono le galassie che si muovono uno spazio-tempo preesistente. Inoltre, non vi \`e un centro privilegiato dell'espansione.
\item
\`E possibile avere una velocit\`a relativa tra le galassie superiore di quella della luce, $v>c$. Questo non viola per\`o il principio di relativit\`a; localmente tutte le velocit\`a relative nello stesso evento spazio-temporale sono subluminali.
\item L'osservazione di uno spostamento verso il rosso dello spettro emesso dalle galassie conferma la recessione delle galassie, e dunque l'espansione dell'universo con $\dot a>0$.
\mynote{Talk here about  Lema\^\i tre?}
\end{enumerate}


\paragraph{Le simmetrie della metrica di Robertson-Walker.} Abbiamo visto che l'omogeneit\`a delle sezioni spaziali $\Si_\tau$ implica l'esistenza di un gruppo di isometrie\index{gruppo di isometrie} che permettono di trasformare un punto qualunque di $\Si_k$ in un qualunque altro punto della stessa sezione spaziale. Queste isometrie si estendono in modo naturale all'intero spazio-tempo di Robertson-Walker.
\begin{Esercizio}\label{es:rwiso}
Mostrare che se $\xi$ \`e un vettore di Killing della metrica tridimensionale $d\sigma^2=h_{ij}\,dx^idx^j$, allora \`e anche un vettore di Killing della metrica quadridimensionale $ds^2=-d\tau^2+a^2(\tau)\,d\sigma^2$. L'isometria dello spazio tridimensionale generata dal vettore di Killing\index{Killing, vettore di}\footnote{Un campo vettoriale $\xi^\mu$ si dice campo vettoriale di Killing se $\nabla_{(\mu}\xi_{\nu)}=0$. Un tale campo vettoriale genera un gruppo a un parametro di isometrie della metrica $g_{\mu\nu}$. \`E un facile esercizio verificare che, se $t^\mu$ \`e il vettore tangente a una geodetica parametrizzata affinemente, $\xi\cdot t$ \`e costante lungo la geodetica.} $\xi$ si estende dunque naturalmente a tutto lo spazio-tempo, e l'isometria risultante \`e generata dallo stesso vettore di Killing $\xi$ definito sull'intera variet\`a quadridimensionale.
\end{Esercizio}
Le sezioni spaziali dello spazio-tempo di Robertson-Walker sono $\mc S^3$, $\mathbb R^3$ o $\mathbb H^3$ a seconda del valore di $k$. Queste metriche a curvatura costante hanno sei vettori di Killing linearmente indipendentemente, che generano rispettivamente i gruppi di isometria\footnote{Il gruppo delle rotazioni dello spazio euclideo in $n+1$ dimensioni \`e il gruppo ortogonale speciale $SO(n)$, e i suoi elementi corrispondono ad isometrie della $n$-sfera\index{$n$-sfera}. Il gruppo euclideo $ISO(n)$ \`e il gruppo di simmetria dello spazio euclideo $n$-dimensionale, ed \`e il prodotto delle rotazioni $SO(n)$ e delle $n$ traslazioni indipendenti. Infine, lo spazio iperbolico\index{spazio!iperbolico} $n$-dimensonale $\mathbb{H}^n$ ha per gruppo di isometrie $SO(n,1)$, che corrisponde al gruppo di Lorentz in $n+1$ dimensioni.
}\index{gruppo di isometrie}
$SO(4)$, $ISO(3)$ e $SO(3,1)$. Genericamente, lo spazio-tempo di Robertson-Walker ha lo stesso gruppo di isometrie. Per forme particolari del fattore di scala $a(\tau)$, il gruppo di simmetria dell'universo di Robertson-Walker pu\`o essere pi\'u esteso. Ad esempio, l'universo di de~Sitter ha gruppo di isometrie $SO(1,4)$, Minkowski ha l'intero gruppo di Poincar\'e $ISO(3,1)$, e anti-de~Sitter ha per gruppo di isometrie il gruppo conforme quadridimensionale $SO(3,2)$.

Queste trasformazioni, in quanto isometrie dell'intero spazio-tempo, trasformano le linee di mondo degli osservatori isotropi in linee mondo di altri osservatori isotropi. Inoltre, visto che gli osservatori isotropi $\mc A$ e $\mc B$ che passano da due punti $x_A^i$ e $x_B^i$ di $\Sigma_\tau$ sono definiti in modo univoco, l'estensione all'intero spazio-tempo dell'isometria spaziale che trasforma $x_A^i$ in $x_B^i$ trasforma la linea di mondo di $\mc A$ nella linea di mondo di $\mc B$.

\paragraph{La propagazione della luce e il redshift cosmologico.}
L'osservazione pi\'u diretta dell'espansione dell'universo proviene dallo spostamento verso il rosso\index{spostamento verso il rosso|see{\emph{redshift}}} (\textit{redshift}\index{redshift}) dello spettro della luce proveniente dalle galassie lontane. In questo paragrafo vogliamo quantificare la differenza tra frequenza al momento di emissione del segnale e frequenza misurata al momento di ricezione, seguendo lungo una geodetica nulla il fotone che connette gli eventi di emissione e di ricezione.

Supponiamo che un osservatore isotropo $\mathcal A$ emetta un fotone di frequenza $\omega_1$ al tempo $\tau_1$ (evento $P_1$) e che un secondo osservatore isotropo $\mathcal B$ ne misuri la frequenza $\omega_2$ al tempo $\tau_2$ (evento $P_2$).
Nell'approssimazione dell'ottica geometrica, il fotone propaga lungo una geodetica nulla -- che chiameremo $\gamma$ -- e la frequenza misurata da un osservatore in moto con quadrivelocit\`a $u^\mu$ \`e
\beq
\omega=-k_\mu u^\mu.
\label{freqdef}\eeq
Qui $k^\mu$ \`e il vettore d'onda del segnale, proporzionale al vettore tangente a $\gamma$ parametrizzata affinemente.

Come visto nel paragrafo precedente, esiste un'isometria dello spazio-tempo che manda la linea di mondo di $\mc A$ in quella di $\mc B$. Sia $\xi$ il campo vettoriale di Killing che la genera (vedi esercizio~\ref{es:rwiso}).
Si mostra che la proiezione su $\Si_\tau$ del vettore tangente a $\gamma$, e quindi di $k^\mu$, \`e allineata al vettore di Killing $\xi$, in ogni istante $\tau_1\leq\tau\leq\tau_2$.
Per semplicit\`a, lo dimostriamo restringendoci al caso di un universo piatto ($k=0$\mynote{notation clash $k$ and $k^\mu$...}). La metrica si scrive in questo caso
\eq
ds^2=-d\tau^2+a^2(\tau)\lp dx^2+dy^2+dz^2\rp.
\eeq
I campi vettoriali $\p_x$, $\p_y$, e $\p_z$ sono campi vettoriali di Killing, e senza perdita di generalit\`a possiamo allineare gli osservatori $\mc A$ e $\mc B$ lungo $x$. L'osservatore $\mc A$ emette dunque il fotone in direzione $x$ e al momento di emissione abbiamo
\eq
\left.k\cdot\p_y\right|_{P_1}=\left.k\cdot\p_z\right|_{P_1}=0.
\eeq
Queste relazioni rimangono valide lungo tutta la curva $\gamma$ visto che la propagazione \`e geodetica e perci\`o la proiezione $k_\perp^\mu$ di $k^\mu$ sulle ipersuperfici spaziali \`e allineata lungo il vettore di Killing $\xi=\p_x$. Argomenti simili si possono fare anche nel caso di un universo chiuso ($k=+1$) e aperto ($k=-1$).

Segue dunque che il vettore d'onda $k^\mu$ si decompone nella somma in una componente lungo $u$ (ortogonale alle ipersuperfici $\Sigma_\tau$), e una componente lungo $\xi$ (tangente alle ipersuperfici $\Sigma_\tau$).
Usando questa decomposizione -- e notando che $\xi^2=a^2(\tau)$ -- scriviamo il vettore d'onda come la seguente combinazione lineare,
\eq
k^\mu=-(k\cdot u)u^\mu+\frac{k\cdot \xi}{\sqrt{\xi^2}}\frac{\xi^\mu}{\sqrt{\xi^2}}.
\eeq
La curva $\gamma$ \`e una geodetica di tipo luce, e dunque 
$k^\mu$ \`e un vettore nullo
\eq
k^2=0=-(k\cdot u)^2+\frac{k\cdot\xi}{\xi^2}.
\eeq
Dalla definizione~\eqref{freqdef}, otteniamo quindi la frequenza del segnale nell'evento $P$, misurata da un osservatore comovente,
\eq
\om_P=\left.\frac{k\cdot\xi}{\sqrt{\xi^2}}\right|_P.
\eeq
Dato che $k\cdot\xi$ \`e costante lungo la geodetica $\ga$, possiamo confrontare la frequenza del segnale al momento di emissione in $P_1$ con quella al momento di ricezione in~$P_2$,
\eq
\frac{\om_2}{\om_1}=\frac{\left.\sqrt{\xi^2}\right|_{P_1}}{\left.\sqrt{\xi^2}\right|_{P_2}}
=\frac{a(\tau_1)}{a(\tau_2)}.
\label{redshift}\eeq
Vediamo che il risultato dipende solamente dal valore iniziale e quello finale del fattore di scala. Mentre l'universo si espande, la lunghezza d'onda di ogni fotone aumenta proporzionalmente all'espansione. A un tempo successivo $\tau_2>\tau_1$, abbiamo $a(\tau_2)>a(\tau_1)$ e dunque la frequenza misurata $\om_2<\om_1$ \`e minore di quella emessa, ovvero la radiazione ha subito uno \textit{spostamento verso il rosso}\index{redshift}\mynote{add a couple of figures for this paragraph}.
Una derivazione alternativa della formula del redshift cosmologico si trova nell'esercizio~\ref{es:redshift}.

\paragraph{Legge di Hubble.}
\`E utile caratterizzare il redshift subito dalla radiazione emessa da una sorgente con la differenza relativa tra la lunghezza d'onda emessa $\la_1$ e la lunghezza d'onda ricevuta $\la_2$. Questa definisce il \textit{parametro di redshift}\index{parametro di redshift}~$z$\index{$z$|see{parametro di redshift}},
\eq
z=\frac{\lambda_2-\lambda_1}{\lambda_1}.
\eeq
Dalla formula~\eqref{redshift} segue dunque che
\eq
z=\frac{a(\tau_2)}{a(\tau_1)}-1.
\label{zdef}\eeq
Se le galassie sono sufficientemente vicine tra loro, possiamo considerare il fattore di scala approssimativamente costante durante la propagazione del segnale. La loro distanza propria \`e dunque $d\approx\tau_2-\tau_1$, e un'espansione di Taylor fornisce $a(\tau_2)=a(\tau_1)+\dot a(\tau_1)(\tau_2-\tau_1)+\cdots$. Perci\`o il parametro di redshift, in prima approssimazione, segue una legge lineare,
\eq\label{leggeHubble}
z=H_0d.
\eeq
Questa \`e la legge di Hubble\index{Hubble!legge di}, una relazione tra redshift e distanza che abbiamo gi\`a incontrato, e che stabilisce evidenza diretta dell'espansione dell'universo.

\begin{figure} 
   \centering
   \begin{tikzpicture}
	\draw[->] (-0.1,  0.0) -- ( 9.0, 0.0) node[right]{$d$}; 
	\draw[->] ( 0.0, -0.1) -- (-0.0, 5.0) node[above]{$z$}; 
	\draw [red,thick] (0.0, 0.0) .. controls (30:4) and (29:5) .. (27:9) node[midway, below, sloped]{\small espansione accelerata};
	\draw [blue, thick, loosely dashed] (0,0)--(30:9.0) node[midway, above, sloped]{\small legge di Hubble};
	\draw [blue, thick, loosely dashed] (0,0)--(30:9.4);
\end{tikzpicture}
\caption{\textbf{Espansione accelerata.} Se il parametro di Hubble \`e costante, il parametro di redshift $z$ \`e una funzione lineare della distanza dalla galassia di provenienza della luce. Per galassie lontane, si misura una deviazione dalla legge lineare, corrispondente a un'espansione accelerata.}
   \label{fig:accelerata}
\end{figure}
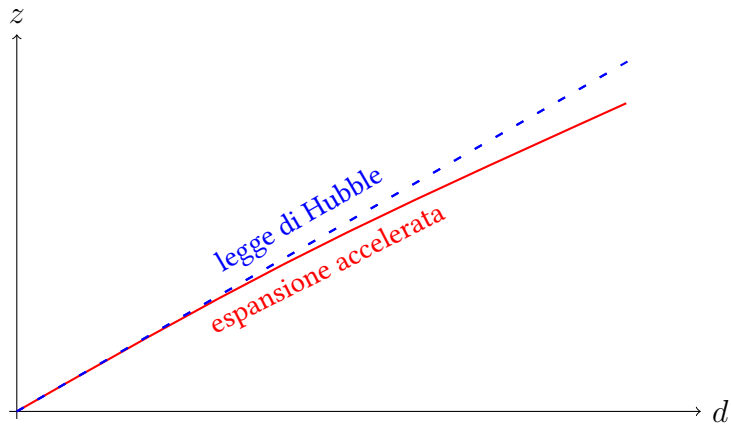

\paragraph{Deviazioni dalla legge di Hubble.}
Se il parametro di Hubble non \`e costante, vi saranno deviazioni dalla linearit\`a nella legge di Hubble~\eqref{leggeHubble}. Misure degli spettri di supernove di tipo Ia distanti (attraverso relazione magnitudo-redshift) mostrano che lo spostamento verso il rosso subito \`e minore di quello previsto dalla legge di Hubble (vedi figura~\ref{fig:accelerata}\mynote{add figure}).
In altre parole, quando la radiazione elettromagnetica \`e stata emessa era $H<H_0$, e dunque l'espansione dell'universo \`e in una fase di accelerazione.\mynote{add discussion of deceleration parameter $q_0$? Exercise with expansion!}

\paragraph{Leggi di conservazione.}
L'universo \`e omogeneo e isotropo solo su grande scala. Sono le osservabili vettoriali e tensoriali mediate su volumi sufficientemente grandi che, per rispettare omogeneit\`a e isotropia, devono potersi esprimere esclusivamente tramite il vettore quadrivelocit\`a~$u^\mu$ e la metrica~$g_{\mu\nu}$.

Ad esempio, si pu\`o definire una corrente numero di galassie, barioni, etc.~che descrive il numero medio di questi oggetti su volumi di scala cosmologica. L'isotropia vincola questa corrente ad essere proporzionale a $u^\mu$. Le sue componenti sono dunque $J^\tau=n(\tau)$ e $J^i=0$, con $n(\tau)$ densit\`a numerica\index{densit\`a numerica} di galassie, barioni etc.
La legge di conservazione del numero $\nabla_\mu J^\mu=\dot n+\Gamma^\mu_{\mu\tau}n=0$ diventa
\eq
\frac{dn}{d\tau}+3\frac{\dot a}{a}n=0,
\eeq
dove abbiamo applicato la comoda formula $\Ga^\mu_{\mu\tau}=\frac{\p}{\p\tau}\ln\sqrt{-g}=3\dot a/a$ per il calcolo della traccia dei simboli di Christoffel che appare nella quadridivergenza.
Questa equazione differenziale \`e separabile e si risolve facilmente. Otteniamo
\eq
n(\tau)=\frac{\textsf{cost.}}{a^3},
\label{numdens}\eeq
ovvero se il numero si conserva, la densit\`a numerica subisce una diluizione con il volume  dell'universo -- proporzionale ad $a^3$ -- durante la sua espansione.

\section{Le equazioni di Friedmann e la dinamica cosmologica}
Vogliamo ora determinare l'evoluzione dell'universo in modo pi\'u quantitativo, ovvero ottenere la forma precisa del fattore di scala $a(\tau)$ risolvendo le equazioni di Einstein $G_\munu=8\pi\newt T_\munu$. Per fare questo, come gi\`a discusso, abbiamo bisogno di conoscere il contenuto in energia e impulso dell'universo, sotto forma del tensore energia-impulso $T_\munu$.

Enunciando il principio cosmologico abbiamo implicitamente ipotizzato che anche la distribuzione di energia e di materia nell'universo deve essere spazialmente omogenea e isotropa. Per rispettare quest'ultima ipotesi, dobbiamo poter esprimere $T_\munu$ come una combinazione lineare degli unici tensori simmetrici di tipo $(0,2)$ che abbiamo a disposizione: $g_\munu$ e $u_\mu u_\nu$. La forma pi\'u generale che pu\`o assumere il tensore energia-impulso di una distribuzione isotropa \`e perci\`o quella del fluido perfetto
\begin{equation}
T_\munu=\rho(x)\,u_\mu u_\nu+P(x)\left(g_\munu+u_\mu u_\nu\right).
\label{Tfluidoperfetto}\end{equation}
Infatti, se cos\'\i\ non fosse, la proiezione di $T_\munu u^\nu$ su $\Sigma_\tau$ definirebbe una direzione privilegiata, in violazione dell'isotropia osservata.
Sono state introdotte due nuove funzioni,
$\rho(x)$ che rappresenta la densit\`a di energia nell'evento $x$, e $P(x)$ che corrisponde alla pressione del fluido in quel punto. 
L'ipotesi di omogeneit\`a spaziale implica invece che la densit\`a di energia $\rho$ e la pressione $P$ devono essere costanti sulle superfici di omogeneit\`a $\Si_\tau$, e in conseguenza possono dipendere solo dalla coordinata temporale $\tau$.

Questo per\`o non \`e sufficiente per risolvere le equazioni di Einstein e studiare la dinamica cosmologica. Infatti, siamo in presenza di tre funzioni incognite $a(\tau)$, $\rho(\tau)$, e $P(\tau)$, ma le equazioni di Einstein forniscono solamente due equazioni indipendenti. Vedremo questo esplicitamente pi\'u avanti, ma si capisce facilmente che non pu\`o essere altrimenti dopo aver imposto l'isotropia spaziale.
Infatti, definendo $\mc E_\munu=G_\munu-8\pi\newt T_\munu$, le equazioni di campo si scrivono $\mc E_\munu=0$.
Ma le componenti fuori diagonale $\mc E_{\tau i}$ si annullano automaticamente: una tale componente non nulla definirebbe in effetti una direzione privilegiata su $\Sigma_\tau$. Per lo stesso motivo, le componenti spaziali delle equazioni del moto devono essere proporzionali alla metrica indotta sulle ipersuperfici di omogeneit\`a, $\mc E_{ij}=\varepsilon h_{ij}$. Rimangono solo due equazioni indipendenti, $\mc E_{\tau\tau}=0$ e $\varepsilon=0$.

Per fare ulteriori progressi nella descrizione della dinamica dell'universo \`e necessario fare qualche ipotesi sulla natura della materia presente nell'universo. Su grande scala la materia contenuta nell'universo pu\`o approssimarsi con un fluido descritto da un'\emph{equazione di stato}
\beq
P=P(\rho).
\label{eqstato}\eeq
Una tale relazione riduce effettivamente a due le incognite del problema, e permette di risolvere completamente le equazioni di campo.
I fluidi perfetti pertinenti alla cosmologia hanno un'equazione di stato lineare,
\beq
P=w\rho,
\label{wstate}\eeq
dove $w$ \`e una costante. I valori rilevanti del parametro $w$ sono i seguenti:
\begin{enumerate}
\item \textsf{La materia} si raggruppa in stelle, galassie, ammassi di galassie...
Su scala cosmica la singola galassia pu\`o venire descritta come un granello di polvere, e per il fatto che le velocit\`a casuali relative tra le galassie  sono piccole, possiamo trascurarne la pressione. L'equazione di stato per la materia \`e dunque semplicemente $P=0$, che si ottiene per parametro $w=0$.
Ulteriori contributi alla densit\`a di energia totale della materia sono dovuti ai gas e polveri interstellari/intergalattici, e alla materia oscura\index{materia oscura} (in questo caso si parla di \emph{cold dark matter}, o CDM). La densit\`a di energia totale dovuta a tutti questi contributi viene indicata con $\rhom$.
\item \textsf{La radiazione}, ad esempio quella di corpo nero, \`e caratterizzata dalla traccia nulla del corrispondente tensore degli sforzi $T^\mu{}_\mu=0$, che implica per il fluido perfetto $P=\rho/3$, ovvero un parametro $w=1/3$. Il contributo preponderante alla densit\`a di energia della radiazione $\rhor$ \`e dovuto alla radiazione cosmica di fondo a 2.7\textsf{K} che permea l'intero universo. Un ulteriore contributo \`e dato dai \emph{neutrini cosmici di fondo} (analogo al CMB). Queste particelle, di massa estremamente piccola, sono in regime ultra-relativistico, e dunque si comportano a tutti gli effetti come un fluido di radiazione.
\item \textsf{L'energia oscura,} \`e necessaria per spiegare l'accelerazione che osserviamo nell'espansione dell'universo. I limiti pi\'u stringenti sulla sua equazione di stato si ottengono combinando i risultati di \textsf{Planck} a quelli dell'osservazioni di supernove ed altri: $w=-1.1\pm0.1$. Il suo contributo \`e compatibile con quello di una costante cosmologica $\Lambda>0$ nelle equazioni di Einstein. Infatti, portata al membro di destra, implica
\beq
G_{ab}=8\pi\newt\lp T_{ab}+T^\Lambda_{ab}\rp,\qquad\text{con}\quad
T^\La_{ab}=-\frac{\La}{8\pi\newt}g_{ab}.
\label{TLambda}\eeq
Dunque, la presenza della costante cosmologica \`e equivalente a quella di un fluido perfetto con equazione di stato \eqref{eqstato} dove $w=-1$. Ad essa \`e associata una densit\`a di energia
\eq
\rho_\La=\frac\La{8\pi\newt}.
\label{rhoLambda}\eeq
Vediamo che $\La>0$ implica $\rho_\Lambda>0$. In questo capitolo identificheremo l'\textit{energia oscura}\index{energia oscura} (dark energy) con questo fluido perfetto con $w=-1$, compatibilmente con le osservazioni. La sua natura per\`o rimane misteriosa ed \`e un importante problema aperto della fisica teorica.
\end{enumerate}
Tutte queste componenti contribuiscono all'energia totale presente nell'universo, $\rho=\rho_\Lambda+\rhom+\rhor$. Tuttavia, la dinamica dell'universo, a seconda della fase della sua storia che attraversa, \`e tipicamente guidata dalla singola componente preponderante di densit\`a di energia.
Parleremo quindi di universo dominato da materia, dominato da radiazione, o dominato dall'energia di vuoto.

\begin{figure} 
   \centering
   \begin{tikzpicture}
	\draw[->] (-.6,0) -- (9.2,0); 
	\draw[->] (-.5,-.1) -- (-.5,6.2); 
\draw [name path=radiation, orange, thick, postaction={decorate,transform shape, decoration={markings, mark=at position .15 with \node[above] {$\rhor$};}}] ({.1},{6.1-4*.1/6})--({9},{6.1-4*9/6});
\draw [name path=matter, red, thick, postaction={decorate,transform shape, decoration={markings, mark=at position .55 with \node[above] {$\rhom$};}}] (.1,{5.5-3*.1/6})--(9,{5.5-3*9/6});
\draw [name path=darkenergy, blue, thick] (-0.5,2.1)--(9,2.1) node[very near end, above]{$\rho_\La$};

\fill [red!40!white, opacity=.1, name intersections={of=matter and darkenergy, by={a}}, name intersections={of=matter and radiation, by={b}}]
    (a |- 0,0) rectangle (b |- 0,6);
\fill [orange!40!white, opacity=.1, name intersections={of=matter and radiation, by={b}}, path fading=fade left]
    (-.5,0) rectangle (b |- 0,6);
\fill [blue!40!white, opacity=.1, name intersections={of=matter and darkenergy, by={a}}, path fading=fade right]
    (a |- 0,0) rectangle (9,6);

\draw [dashed, name intersections={of=matter and darkenergy, by={a}}, xshift=10cm]
    ([xshift=0.2cm]a |- 0,0) node[anchor=north,black]{$0$} -- ([xshift=0.2cm]a |- 0,4.5) node[anchor=north west] {\small oggi};
\draw [help lines, name intersections={of=matter and darkenergy, by={a}}]
    (a |- 0,0) -- (a |- 0,6) node[anchor=north west] {\small $\La$ dominated};
\draw [help lines, name intersections={of=matter and radiation, by={b}}]
    (b |- 0,0) node[anchor=north,black]{$\ln(a_{\textsf{eq}}/a_0)$} -- (b |- 0,6) node[anchor=north west] {\small matter dominated};
\draw (0,3.1) node[help lines, anchor=north west] {\small radiation dominated};
	\draw (9.2,0) node[right] {$\ln\lp a/a_0\rp$};
	\draw (-.5,6.2) node[above] {$\ln\lp\rho_{\mathsf{i}}/\rho_0\rp$};
\end{tikzpicture}
   \caption{\textbf{Evoluzione della composizione dell'universo.} }
   \label{fig:comp}
\end{figure}
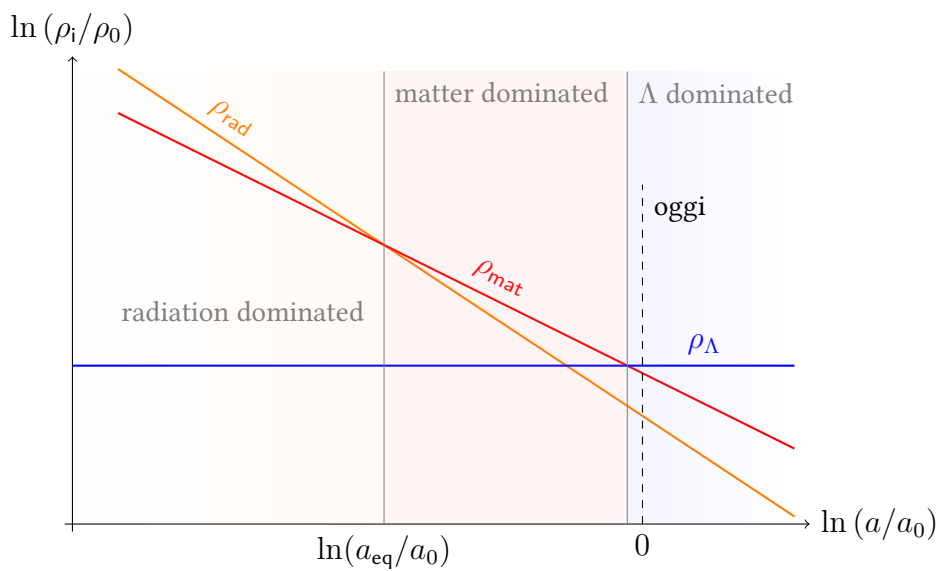

\paragraph{Evoluzione della densit\`a di energia.}
L'evoluzione di $\rho(\tau)$ si ottiene facilmente dalla componente temporale dell'equazione di conservazione del tensore energia-impulso, $\nabla_\mu T^\mu{}_\tau=0$. Per valutare la derivata covariante abbiamo bisogno dei simboli di Christoffel per la metrica di Robertson-Walker.
\begin{Esercizio}
Verificare che i simboli di Christoffel non nulli della la metrica di Robertson-Walker \eqref{RWmetric} sono dati da:
\begin{align}
\Ga^\tau_{\chi\chi}&=a\dot a,
&\Ga^\tau_{\theta\theta}&=a\dot aS_k^2,
&\Ga^\tau_{\phi\phi}&=a\dot aS_k^2\sin^2\theta,\vphantom{\frac{\dot a}a}\nonumber\\
\Ga^\chi_{\theta\theta}&=-S_kS_k',
&\Ga^\chi_{\phi\phi}&=-S_kS_k'\sin^2\theta,
&\Ga^\theta_{\phi\phi}&=-\sin\theta\cos\theta,\nonumber\\
\Ga^\chi_{\chi\tau}&=\Ga^\theta_{\theta\tau}=\Ga^\phi_{\phi\tau}=\frac{\dot a}a,
&\Ga^\theta_{\theta\chi}&=\Ga^\phi_{\phi\chi}=\frac{S_k'}{S_k},
&\Ga^\phi_{\phi\theta}&=\cot\theta.
\end{align}
\textbf{Nota:} Sia $\sigma_{ij}=h_{ij}/a^2(\tau)$ la metrica della sfera/piano/iperboloide unitario.
Allora possiamo scrivere alcuni simboli di Christoffel in forma pi\'u compatta,
\eq
\Ga^\tau_{ij}=a\dot a\sigma_{ij},\quad
\Ga^i_{j\tau}=\frac{\dot a}a\delta^i{}_j.
\eeq
\end{Esercizio}
Otteniamo cos\'\i\ l'equazione di conservazione dell'energia per il fluido perfetto,
\beq
\dot\rho=-3(\rho+P)\frac{\dot a}a,
\label{econs}\eeq
che si pu\`o anche riscrivere nella forma pi\'u familiare
\eq
\frac{d}{d\tau}\lp\rho a^3\rp=-P\frac{d}{d\tau}\lp a^3\rp.
\eeq
La legge di conservazione dell'energia per il tensore energia-impulso del fluido non \`e altro che la prima legge della termodinamica applicata localmente al fluido perfetto: la variazione dell'energia in un dato elemento di fluido \`e data dal lavoro effettuato dalle forze di pressione.
Se il fluido verifica l'equazione di stato \eqref{wstate}, la conservazione dell'energia \`e dettata da un'equazione differenziale separabile e si pu\`o risolvere con metodi elementari. La soluzione lega la densit\`a di energia al fattore di scala,
\eq
\rho(\tau)=\rho_0\lp\frac{a_0}{a(\tau)}\rp^{3(1+w)}
\label{rho(a)}\eeq 
con $\rho_0$ costante d'integrazione che determina il valore attuale della densit\`a di energia. Il comportamento dei fluidi di rilevanza cosmologica \`e il seguente:
\begin{enumerate}
\item\textsf{Materia non relativistica ($w=0$)}\index{materia non relativistica}\index{polvere}
\quad Abbiamo un fluido di polvere ($P=0$), la cui densit\`a di energia \`e determinata dalla massa a riposo delle sue particelle costitutive, ovvero \`e proporzionale alla densit\`a numerica di queste particelle. In conseguenza, la densit\`a di energia diminuisce a causa della diluizione di queste particelle durante l'espansione dell'universo, $\rhom\propto1/a^3$, in accordo con quanto trovato in \eqref{numdens}.
\item\textsf{Radiazione ($w=1/3$)}\index{radiazione}\mynote{materia ultra-relativistica} \quad In questo caso abbiamo un diminuzione pi\'u rapida della densit\`a di energia, $\rhor\propto1/a^4$. Un fattore $1/a^3$ proviene dalla diminuzione della densit\`a numerica di fotoni dovuta all'espansione dell'elemento di volume, analogamente a quanto accade nel caso della polvere. Vi \`e un'ulteriore perdita di energia $\propto1/a$ causata dal redshift cosmologico, in quanto la lunghezza d'onda dei singoli fotoni aumenta proporzionalmente al fattore di scala $a$.
Questa equazione di stato vale anche per \emph{materia ultra-relativistica},\index{materia ultra-relativistica|see{radiazione}}
come ad esempio i neutrini.
\item\textsf{Energia oscura ($w=-1$)} \quad In questo caso la densit\`a di energia $\rho_\Lambda$ rimane costante, indipendentemente dalla variazione del fattore di scala.
\end{enumerate}
Vediamo cos\'\i\ che $\rho_{\textsf{rad}}$ decresce pi\'u rapidamente che $\rho_{\textsf{mat}}$, mentre $\rho_\Lambda$ rimane costante. Attualmente stiamo entrando in una fase dominata dall'energia oscura, ma nel passato -- e durante la maggior parte della sua vita -- l'universo era dominato dalla materia. Solo nei primi istanti dopo il Big Bang
l'universo era dominato dalla radiazione (vedi figura~\ref{fig:comp}). L'epoca di uguaglianza materia-radiazione, il periodo con $\rhom=\rhor$ durante il quale la dominanza \`e passata alla materia, \`e avvenuta a redshift $z_{\textsf{eq}}=3360\pm70$\index{$z_{\mathsf{eq}}$}, quando l'et\`a dell'universo era di circa 61\,000~anni.\label{pag:zeq}

\paragraph{Equazioni di Friedmann.} Veniamo finalmente alla forma delle equazioni di Einstein per un universo omogeneo ed isotropo. Per questo dobbiamo anzitutto calcolare la curvatura (e in particolare il tensori di Einstein) per la metrica di Robertson-Walker. Questo \`e l'oggetto del prossimo esercizio.
\begin{Esercizio}
Verificare che, per la metrica di Robertson-Walker \eqref{RWmetric}, le componenti non nulle del tensore di Riemann sono date da
\begin{align}
&R_{\tau\chi\tau}{}^\chi=R_{\tau\theta\tau}{}^\theta=R_{\tau\phi\tau}{}^\phi=-\frac{\ddot a}a,\\
&R_{\chi\theta\chi}{}^\theta=R_{\chi\phi\chi}{}^\phi=\dot a^2+k,
&&R_{\theta\phi\theta}{}^\phi=\lp\dot a^2+k\rp\Si_k^2,
\end{align}
quelle del tensore di Ricci da
\begin{align}
&R_{\tau\tau}=-\frac{3\ddot a}a,
&&R_{\chi\chi}=a\ddot a+2\lp\dot a^2+k\rp&\\
&R_{\theta\theta}=R_{\chi\chi}\Si_k^2,
&&R_{\phi\phi}=R_{\chi\chi}\Si_k^2\sin^2\theta,&
\end{align}
e che in conseguenza lo scalare di curvatura \`e dato da
\eq
R=6\lp\frac{\ddot a}{a}+\frac{\dot a^2+k}{a^2}\rp.
\eeq
Ottenere infine il tensore di Einstein, le cui componenti non nulle sono date da
\begin{align}
&G_{\tau\tau}=\frac{3}{a^2}\lp\dot a^2+k\rp,
&&G_{\chi\chi}=-2a\ddot a-\lp\dot a^2+k\rp&\\
&G_{\theta\theta}=G_{\chi\chi}\Si_k^2,
&&G_{\phi\phi}=G_{\chi\chi}\Si_k^2\sin^2\theta.&
\end{align}
\end{Esercizio}
Utilizzando il risultato dell'esercizio precedente, troviamo che le equazioni di Einstein $G_\munu=8\pi\newt T_\munu$, in presenza di un fluido perfetto, si riducono al sistema di equazioni
\bseq
\begin{align}
\lp\frac{\dot a}a\right)^2&=\frac{8\pi\newt}3\rho-\frac k{a^2},
\label{F1}\\
\frac{\ddot a}a&=-\frac{4\pi\newt}3(\rho+3P).
\label{F2}
\end{align}
\eseq
Queste due equazioni dettano l'evoluzione dell'universo e sono dette \textit{equazioni di Friedmann}\index{equazioni!di Friedmann}. Come abbiamo visto, per risolverle vanno completate con l'aggiunta dell'equazione di stato $P=P(\rho)$ per il fluido perfetto. \`E importante osservare che le equazioni di Friedmann~\eqref{F1} e~\eqref{F2} implicano la legge di conservazione dell'energia \eqref{econs}. Questo non \`e sorprendente, visto che quest'ultima \`e una conseguenza dell'equazione di Bianchi contratta.

L'universo contiene tuttavia pi\'u componenti di fluido perfetto, una per l'energia oscura, una per la materia e una per la radiazione e particelle ultra-relativistiche. Nel membro di destra delle equazioni di Einstein va messo dunque il tensore energia-impulso totale, somma di queste componenti. Segue che la densit\`a $\rho$ e la pressione $P$ che appaiono nelle equazioni di Friedmann~\eqref{F1} e~\eqref{F2} sono quelle totali, $\rho=\rho_\Lambda+\rhom+\rhor$ e $P=P_\Lambda+P_{\textsf{rad}}$. Abbiamo ora ulteriori funzioni incognite, ma faremo l'ipotesi che le varie componenti di fluido non interagiscono tra loro, se non tramite l'interazione gravitazionale. Le singole componenti del fluido cosmologico verificano dunque l'equazione di conservazione~\eqref{econs}, e dunque ogni componente di densit\`a di energia verifica la legge~\eqref{rho(a)} con l'appropriato valore di $w$.

Motivati dalla forma assunta dalla prima equazione di Friedmann~\eqref{F1}, enunciamo la seguente definizione.
\definizione{Si definisce il \textit{parametro di densit\`a}\index{parametro di densit\`a}
\eq
\Omega=\frac{8\pi\newt}{3H^2}\rho.
\eeq
Si pu\`o scrivere $\Omega=\rho/\rhocrit$, dove abbiamo definito la \textit{densit\`a critica}\index{densit\`a critica}
\eq
\rhocrit=\frac{3H^2}{8\pi\newt}.
\eeq
}
Con questa definizione, la prima equazione di Friedmann~\eqref{F1} si riscrive
\eq
\Omega-1=\frac k{H^2a^2}.
\label{F1om}\eeq
Indipendentemente dalle ipotesi che avanziamo sui costituenti dell'universo, il segno di $\Omega-1$ determina $k$, e dunque la natura delle sezioni spaziali in cui viviamo,
\begin{enumerate}
\item Se $\rho_0<\rhocritzero$, allora $\Om_0<1$ e $k=-1$, e l'universo \`e \textit{aperto}\index{universo!aperto},
\item Se $\rho_0=\rhocritzero$, allora $\Om_0=1$ e $k=0$, e l'universo \`e \textit{piatto}\index{universo!piatto},
\item Se $\rho_0>\rhocritzero$, allora $\Om_0>1$ e $k=+1$, e l'universo \`e \textit{chiuso\index{universo!chiuso}}.
\end{enumerate}
Le osservazioni ci forniscono un valore attuale del parametro di densit\`a $\Omega_0=1.000(7)$; il nostro universo \`e molto vicino a un'universo piatto, ma la precisione degli esperimenti non ci permette di determinare se la densit\`a di energia \`e maggiore o inferiore a quella critica. Torneremo su questo punto pi\'u avanti.

Il parametro densit\`a totale si decompone nella somma dei contributi dei vari fluidi relativistici presenti nell'universo,
$\Om=\Om_{\Lambda}+\Om_{\mathsf{mat}}+\Om_{\mathsf{rad}}$. Il contributo di materia senza pressione (polvere) $\Om_{\mathsf{mat}}$ consiste in una componente di materia barionica $\Om_{\mathsf{b}}$, e una di materia oscura $\Om_{\mathsf{CMB}}$. La componente di radiazione $\Om_{\mathsf{rad}}$ raccoglie il contributo delle particelle relativistiche, i fotoni $\Om_{\mathsf{\gamma}}$ provenienti principalmente dalla radiazione cosmica di fondo, e i neutrini $\Om_{\nu}$, la cui massa \`e trascurabile. Infine, il resto della densit\`a di energia presente nell'universo \`e dovuto all'energia oscura $\Om_\La$. I valori osservati per i parametri densit\`a sono riportati nella tabella~\ref{tab:density}. Il modello cosmologico, al quale ci restringiamo in queste lezioni, in cui l'energia oscura proviene da una costante cosmologica, e la materia oscura si comporta come polvere (materia oscura fredda,
\index{materia oscura}
o \emph{cold dark matter}, CDM) si dice \textit{modello $\Lambda$CDM}.
\index{$\Lambda$CDM (modello)}
Ad oggi, \`e il modello pi\'u semplice che permette di spiegare le osservazioni cosmologiche.
\begin{table}
\caption{\textbf{Densit\`a delle varie componenti del fluido cosmologico \cite{Beringer:1900zz}}.}
\label{tab:density}
\centering{\begin{tabularx}{\textwidth}{l X X}
\toprule
Componente & Simbolo & Valore attuale\\
\midrule
\smallskip
Energia oscura & $\Omega_\Lambda$ & $0.685^{+0.017}_{-0.016}$\\
Materia, di cui & $\Omega_{\textsf{mat}}=\Omega_{\textsf{b}}+\Omega_{\textsf{CDM}}$ & $0.315^{+0.016}_{-0.017}$\\
\qquad barioni & $\Omega_{\textsf{b}}$ & $0.0499(22)$\\
\smallskip
\qquad materia oscura (CDM) & $\Omega_{\textsf{CDM}}$ & $0.265(11)$\\
Radiazione, di cui & $\Om_{\mathsf{rad}}$ & $\approx 10^{-4}$\\
\qquad fotoni (CMB)& $\Om_{\gamma}$ & $5.46(19)\times10^{-5}$\\
\qquad neutrini & $\Om_{\nu}$ & $<0.0055$\\
\bottomrule
\end{tabularx}}
\end{table}
\begin{Esercizio}
\textbf{Epoca di uguaglianza materia-radiazione} \quad
Calcolare il parametro di redshift $z_{\mathsf{eq}}$\index{$z_{\mathsf{eq}}$} corrispondente l'epoca dell'uguaglianza ma\-te\-ria-ra\-dia\-zione, considerando solo il contributo della CMB per la radiazione, e confrontare con il valore citato a pagina~\pageref{pag:zeq}. Come cambia questo valore tenendo conto del contributo dei neutrini alla densit\`a di radiazione?
\end{Esercizio}
\begin{Esercizio}
\textbf{Epoca di uguaglianza materia-energia oscura} \quad
Calcolare il parametro di redshift corrispondente all'epoca di uguaglianza tra materia ed energia oscura, e il parametro di redshift quando \`e avvenuto il passaggio tra la fase di espansione decelerata a quella di espansione accelerata. La transizione a un'evoluzione dominata dall'energia oscura \`e avvenuta circa 10 miliardi di anni dopo il Big Bang.
\end{Esercizio}
\section{Evoluzione e fato del nostro universo
}
La materia ordinaria\index{materia ordinaria} soddisfa le condizioni $\rho>0$ e $P\geq0$. Nel nostro universo, osserviamo due componenti di questo tipo, una dovuta alla materia non-relativistica $\rhom$ (polvere\index{polvere}, con $w=0$), e una dovuta alla radiazione\index{radiazione} $\rhor$ (materia ultra-relativistica, principalmente fotoni, ma anche neutrini, con $w=1/3$). Inoltre, abbiamo visto che dobbiamo includere una terza componente di energia oscura, la cui densit\`a $\rho_\La$ soddisfa l'equazione di stato $\rho_\La=-P$ (costante cosmologica, con $w=-1$). Se trascuriamo l'interazione tra le varie componenti di energia, segue che devono conservarsi tutte e tre indipendentemente e dunque, per la \eqref{rho(a)},
\eq
\rhom(a)=\rho_{\textsf{mat},0}\lp\frac{a_0}{a}\rp^3,\quad
\rhor(a)=\rho_{\textsf{rad},0}\lp\frac{a_0}{a}\rp^4,\quad
\rho_{\Lambda}(a)=\rho_{\La,0},
\eeq
con $\rho_{\textsf{mat},0}$, $\rho_{\textsf{rad},0}$, e $\rho_{\La,0}$ i valori attuali della densit\`a di materia, di radiazione, e di energia oscura.
Sostituendo nella prima equazione di Friedmann~\eqref{F1}, troviamo che la dinamica cosmologica \`e descritta dall'equazione
\bseq
\begin{align}
&\frac12\dot a ^2+V(a)=-\frac k2,\label{mech}\\
&V(a)=-\frac{4\pi\newt}3\,a^2\left[\rho_\La+\rhom{}_{,0}\lp\frac{a_0}a\rp^3+\rhor{}_{,0}\lp\frac{a_0}a\rp^4\right].
\label{cosmicV}\end{align}
\eseq
La seconda equazione di Friedmann~\eqref{F2} \`e soddisfatta automaticamente (vedete perch\'e?). L'evoluzione del fattore di scala \`e dunque analoga al moto di una particella classica di massa unitaria e energia $-k/2$ nel potenziale $V(a)$.
Queste equazioni possono essere risolte esattamente con $\Lambda=0$ per universi contenenti sola polvere o sola radiazione (vedi tabella~\ref{tab:scale}). Nel caso generale possiamo invece studiare qualitativamente il comportamento delle soluzioni.
\begin{table}
\caption{\textbf{Soluzione delle equazioni di Friedmann per un universo di polvere e un universo di radiazione ($\Lambda=0$).} Le costanti di integrazione si ottengono da $c_{\textsf{p}}=8\pi\rho_{\textsf{mat}}a^3/3$ e $c_{\textsf{r}}=8\pi\rho_{\textsf{rad}}a^4/3$.}
\label{tab:scale}
{\centering
\begin{tabularx}{1\textwidth}
{l l l l}
\toprule
& $k=+1$ &
$a(\tau)=\frac12c_{\textsf{p}}(1-\cos\eta)$ &
$\tau=\frac12c_{\textsf{p}}\lp\eta-\sin\eta\rp$\\
Polvere &
$k=0$ &
$a(\tau)=\lp\frac{9c_{\textsf{p}}}4\rp^{\frac13}\tau^{\frac23}$ &
\\
& $k=-1$ &
$a(\tau)=\frac12c_{\textsf{p}}(\cosh\eta-1)$ &
$\tau=\frac12c_{\textsf{p}}\lp\sinh\eta-\eta\rp$\\
\midrule
& $k=+1$ &
$a(\tau)=\sqrt{c_{\textsf{r}}}\sqrt{1-\lp1-\frac{\tau}{\sqrt{c_{\textsf{r}}}}\rp^2}$ &
\\
Radiazione &
$k=0$ &
$a(\tau)=\lp4c_{\textsf{r}}\rp^{\frac14}\sqrt{\tau}$ &
\\
& $k=-1$ &
$a(\tau)=\sqrt{c_{\textsf{r}}}\sqrt{\lp1+\frac{\tau}{\sqrt{c_{\textsf{r}}}}\rp^2-1}$ &
\\
\bottomrule
\end{tabularx}}
\label{tab:a}
\end{table}

\paragraph{L'universo statico.} Se il fattore di scala \`e costante, $a(\tau)=a_0$, deve essere $\ddot a=0$ e la seconda equazione di Friedmann~\eqref{F2} richiede $\rho+3P=0$. Questa condizione \`e impossibile da verificare in presenza di sola materia ordinaria. Una costante cosmologica positiva pu\`o tuttavia contribuire con una pressione del vuoto negativa e compensare il contributo della materia ordinaria in modo da soddisfare la condizione di equilibrio. Questo accade per
\eq
\rho_\La=\frac12\rhom+\rhor,
\label{staticLambda}\eeq
nel qual caso abbiamo una soluzione statica.\footnote{Equivalentemente, la condizione~\eqref{staticLambda} si ottiene richiedendo un punto di equilibrio del potenziale~\eqref{cosmicV} per il valore $a_0$ del fattore di scala, i.e.~$V'(a_0)=0$.}
La prima equazione di Friedmann~\eqref{F1} impone poi $k\geq0$. Il caso di universo piatto, con $k=0$, implica $\rho=0$, lo spazio-tempo \`e quello di Minkowski. Per avere una soluzione statica non banale dobbiamo considerare un universo chiuso, con $k=+1$. Lo spazio-tempo \`e in questo caso il prodotto diretto $\mathbb R\times\sphere^3$ con metrica
\eq
ds^2=-d\tau^2+a_0^2\,d\Om^2_3,
\eeq
dove indichiamo con $d\Om^2_3$ metrica unitaria sulla 3-sfera
\footnote
{
Ad esempio, nelle coordinate $(\chi,\theta,\phi)$ che abbiamo definito studiando gli spazi a curvatura costante, la metrica unitaria su $\sphere^3$ assume la forma
$d\Omega_3^2=d\chi^2+\sin^2\chi\lp d\chi^2+\sin^2\theta d\phi^2\rp$.
}
\index{3-sfera}. Questo \`e l'\textit{universo statico di Einstein}\index{universo!statico di Einstein}.
Per avere una soluzione dobbiamo imporre inoltre l'equazione~\eqref{mech} che da
\eq
\rho_\Lambda+\rhom+\rhor=\frac3{8\pi\newt}\frac1{a_0^2}.
\label{esu2}\eeq
Le relazioni~\eqref{staticLambda} e~\eqref{esu2} legano la costante cosmologica e il raggio $a_0$ dell'universo al suo contenuto in radiazione e materia.
\begin{Esercizio}
\textbf{L'universo statico di Einstein}\quad
Mostrare che in presenza di sola materia, l'universo statico di Einstein con costante cosmologica $\Lambda$ deve avere raggio densit\`a di energia della materia dati da
\eq
a_0=\frac1{\sqrt\Lambda},\qquad
\rhom=2\rho_\Lambda=\frac{\Lambda}{4\pi\newt}.
\eeq
Se invece l'universo \`e riempito da sola radiazione,
\eq
a_0=\sqrt{\frac3{2\Lambda}},\qquad
\rhor=\rho_\Lambda=\frac{\Lambda}{8\pi\newt}.
\eeq
\end{Esercizio}
Tuttavia, \`e importante osservare che questa soluzione statica \`e instabile,
\index{universo statico di Einstein!stabilit\`a}
in quanto corrisponde a un \emph{massimo} del potenziale $V(a)$: una qualsiasi perturbazione la porta fuori dall'equilibrio. La dipendenza del fattore di scala dal tempo \`e dunque inevitabile e \emph{la relativit\`a generale ci dice che l'universo in cui viviamo non pu\`o essere statico; l'universo \`e in espansione (o contrazione) e la sua geometria \`e dinamica!}

\paragraph{Espansione accelerata e universo di de~Sitter.}
La seconda equazione di Friedmann~\eqref{F2} implica che, per materia ordinaria, $\ddot a<0$, ovvero la velocit\`a di espansione dell'universo diminuisce nel tempo. Le osservazioni mostrano invece che l'espansione \`e accelerata, e dunque vi deve essere una componente del fluido cosmologico con $w<-1/3$. Abbiamo visto che questa componente -- l'energia oscura -- si comporta come un fluido perfetto con $w=-1$, che corrisponde a una costante cosmologica positiva, vedi le equazioni~\eqref{TLambda}-\eqref{rhoLambda}. Con $k=0$, e se non vi \`e altra materia nell'universo, la prima equazione di Friedmann~\eqref{F1} implica che il parametro di Hubble rimane costante, e dunque che il fattore di scala cresce esponenzialmente,
\eq
a(\tau)=a_0\,e^{H_0\tau},\qquad
H_0=\sqrt{\frac\La3}.
\label{dSscale}\eeq
La metrica risultante \`e quella dell'universo di de~Sitter\index{de~Sitter}
(\emph{dS}).

\paragraph{L'evoluzione passata dell'universo e il Big Bang.}
Il redshift cosmologico che misuriamo suggerisce un universo in espansione, con $\dot a_0>0$. Estrapolando nel passato, e supponendo una velocit\`a di espansione $\dot a$ costante, vediamo che esiste un tempo $\tau_i<\tau_0$ per cui il fattore di scala si annulla, $a(\tau_i)=0$. In quel momento, la distanza propria tra tutti i punti di $\Si_{i}$ \`e nulla. Al tempo $\tau_i$ abbiamo dunque uno stato singolare, in cui la densit\`a di energia e gli invarianti di curvatura divergono. Questo evento iniziale \`e chiamato \textit{Big Bang}\index{Big Bang}, e l'et\`a dell'universo \`e data dal tempo di Hubble\index{Hubble!tempo di} $H_0^{-1}$, circa 14.5
miliardi di anni (vedi figura~\ref{fig:age}). Tuttavia, in un universo formato da materia ordinaria, l'evoluzione \`e decelerata e quindi l'espansione era pi\'u veloce nel passato. L'et\`a dell'universo risulta in questo caso inferiore a quella stimata da $H_0^{-1}$. Stiamo attualmente entrando in una fase di espansione accelerata, dominata dall'energia oscura, ma durante la maggior parte della sua esistenza il nostro universo \`e stato dominato dalla materia. Tenendo conto dei dettagli della sua evoluzione, la miglior stima dell'et\`a dell'universo\index{universo!et\`a dell'} \`e di 13.798$\pm$0.037 miliardi di anni \cite{Beringer:1900zz}.

\begin{figure} 
   \centering
   \begin{tikzpicture}
	\draw[->] (-1.1,0) -- (9,0); 
	\draw[->] (-1,-.1) -- (-1,5); 
	\draw[gray,very thin] (-1,0 |- 30:8)--(30:8);
	\draw[help lines] (30:8 |- 0,0)--(30:8);
	\draw (-1,0 |- 30:8) node[left]{$a_0$};
	\draw (30:8 |- 0,0) node[below]{$\tau_0$};
	\draw [blue,thick] (0,0)--(30:8);
	\draw [blue,thin] (30:8)--(30:9);
	\draw [red,thick] (3,0) .. controls (3,1.5) and (30:5.5) .. (30:8) node[midway, right]{\small universo di polvere};
	\filldraw [gray] (0,0) circle (2pt)
				(30:8) circle (2pt);
	\filldraw [red] (3,0) circle (2pt);
	\draw [shift={(30:8 |- 0,0)}] (0,-1pt) -- (0,1pt);
	\draw [shift={(-1,0 |- 30:8)}] (-1pt,0)--(1pt,0);
	\draw (-1,0 |- 30:8) node[left]{$a_0$};
	\draw (30:8 |- 0,0) node[below]{$\tau_0$};
	\draw (0,0) node[below]{$\tau_i$};
	\draw (9,0) node[right]{$\tau$};
	\draw (-1,5) node[above]{$a(\tau)$};
	\draw[decorate,decoration=brace,shift={(0,-.5)},gray,thick] (30:8 |- 0,0)--(0,0) node[midway, below, black]{\small et\`a di Hubble $H_0^{-1}$};
	\draw[decorate,decoration=brace,shift={(0,.15)},gray,thick] (3,0)--(30:8 |- 0,0) node[midway, above, black]{\small et\`a reale};
\end{tikzpicture}
   \caption{\textbf{Et\`a dell'universo.} In prima approssimazione, $H_0^{-1}$ (et\`a di Hubble) \`e il tempo trascorso dal Big Bang. Per un universo in espansione decelerata (come per l'universo di polvere in figura) l'et\`a reale \`e inferiore. Per un universo in espansione accelerata (come quello in cui viviamo) l'et\`a reale dell'universo \`e maggiore.}
   \label{fig:age}
\end{figure}

\`E opportuno soffermarci per alcune osservazioni sul Big Bang.
\;(a)~Ripercorrendo all'indietro il tempo, abbiamo una contrazione omogenea delle sezioni spaziali $\Sigma_\tau$ fino a raggiungere dimensioni nulle. Questo \`e molto diverso da un universo creato in esplosione: non vi \`e un `centro' dell'espansione, e non c'\`e nemmeno uno spazio o un tempo preesistente al Big Bang.
\;(b)~Bench\'e l'evoluzione del nostro universo sia oggi controllata dall'energia di vuoto, abbiamo visto che nel passato la componente di materia ordinaria (radiazione ai primordi, materia poi) era dominante (vedi figura~\ref{fig:comp}), e la conclusione rimane valida: il Big Bang \`e inevitabile.
\;(c)~I teoremi di singolarit\`a\index{Big Bang!singolarit\`a} (dovuti principalmente a Hawking e Penrose, vedi \cite{Hawking:1973uf}) mostrano che la presenza di una singolarit\`a iniziale non dipende dalle ipotesi di isotropia e omogeneit\`a che abbiamo fatto, ma \`e generica. Senza entrare in dettagli tecnici, questi teoremi ci dicono che un qualunque un universo che \`e in espansione in un determinato istante, e per cui $\rho>0$ e $\press\geq0$, ha una singolarit\`a nel passato.
\;(d)~Infine, \`e importante osservare che, per $a\to0$, sia la densit\`a di energia che la curvatura divergono, e perci\`o effetti gli quantistici diventano via via pi\'u importanti. Per esplorare i primi istanti dopo il Big Bang \`e necessaria una teoria consistente di gravit\`a quantistica (ad esempio la teoria di stringhe).

\paragraph{Evoluzione futura dell'universo}
Per $k=0$ e $k=-1$, la prima equazione di Friedmann \eqref{F1} mostra che
la velocit\`a di espansione $\dot a$ non pu\`o annullarsi se $\rho>0$: un universo piatto o aperto che \`e in espansione a un certo istante rimane in espansione durante tutta la sua evoluzione.\footnote{Si pu\`o vedere che asintoticamente, in presenza di sola materia ordinaria, per $\tau\to+\infty$, $\dot a$ tende a $1$ per l'universo aperto e a $0$ per l'universo piatto (vedi per esempio~\S5.2 di~\cite{Wald:1984rg}).}

Analizziamo in dettaglio il caso $k=0$. Contrariamente a quanto accade durante i primi istanti dell'evoluzione dell'universo, che sono guidati dalla componente di fluido con $w$ massimo (la radiazione), il suo destino finale \`e determinato dalla componente di fluido con $w$ minimo. L'evoluzione tarda \`e dunque dominata dall'energia di vuoto. Se questa \`e assimilabile a una costante cosmologica ($w=-1$), il fattore di scala evolve esponenzialmente secondo l'equazione~\eqref{dSscale}. Le equazioni di Friedmann si integrano facilmente anche nel caso
$w\neq-1$,
\eq
a(\tau)\propto\lp(1+w)(\tau-\bar\tau)\rp^{\frac2{3(1+w)}},
\eeq
con $\bar\tau$ costante di integrazione.
Si trova dunque un'espansione polinomiale nel tempo del fattore di scala quando $w>-1$. Invece, quando $w<-1$, deve essere $\tau<\bar\tau$ in quanto il fattore di scala diverge per $\tau\to\bar\tau$. L'evoluzione termina in un tempo finito con una diluizione infinita di tutta la materia e radiazione, in un evento chiamato \textit{Big Rip}\index{Big Rip}. La conclusione non cambia se $k=-1$, in quanto il termine di curvatura, come i contributi di materia e radiazione, diventa trascurabile nell'evoluzione futura asintotica.

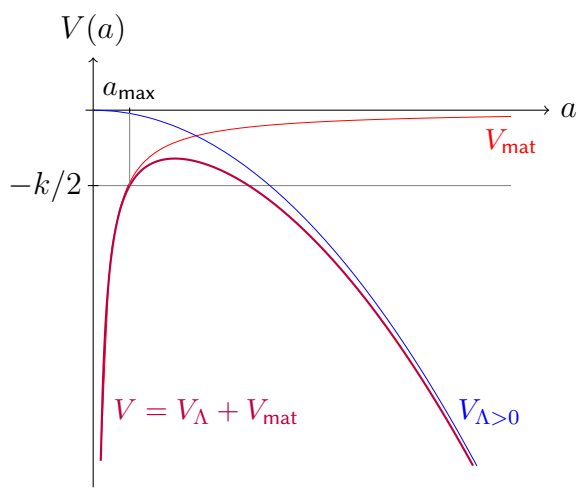
\begin{figure} 
   \centering
\begin{tikzpicture} 
  \draw[->] (-.1,0) -- (6,0) node[right] {$a$};
  \draw[->] (0,-5) -- (0,.7) node[above] {$V(a)$};
  \draw[name path=k, help lines] (0,-1)--(5.5,-1);
\draw[help lines, red, domain=.1:5.5, samples=50, smooth, variable=\t] plot ({\t},{-(.23*8/\t)/4});
\draw (5.5,-.1) node[red, anchor=north] {\small $V_{\textsf{mat}}$};
\draw[help lines, blue, domain=0:5.05,samples=20,smooth, variable=\t] plot ({\t},{-(.74)*\t*\t/4});
\draw (4.65,-4) node[blue, anchor=west] {\small $V_{\La>0}$};
\draw[name path=V, domain=.1:5,samples=200,smooth, variable=\t, purple, thick] plot ({\t},{-(.74+.23*8/(\t*\t*\t))*\t*\t/4});
\draw (3.4pt,-4) node[purple, anchor=west] {\small $V=V_{\La}+V_{\textsf{mat}}
$};
\draw (-2pt,-1)--(2pt,-1);
\draw (0,-1) node[anchor=east] {$-k/2$};
\draw[name intersections={of=k and V}, help lines] (intersection-1) -- +(0,1) node[black, anchor=south]{\small $a_{\textsf{max}}$};
\draw[name intersections={of=k and V}] ($(intersection-1)+(0,1)-(0,1pt)$)--($(intersection-1)+(0,1)+(0,1pt)$);
\end{tikzpicture}
\caption{\textbf{Potenziale cosmologico.} Abbiamo trascurato il contributo della radiazione, che non cambia l'andamento qualitativo del grafico ed \`e trascurabile al di fuori dei primissimi istanti dopo il Big Bang. Se il valore massimo $V_{\textsf{max}}$ del potenziale supera $-1/2$, con $k=+1$, quando $\dot a=0$ l'evoluzione si inverte e inizia una fase di collasso.}
   \label{fig:potcosmo}
\end{figure}

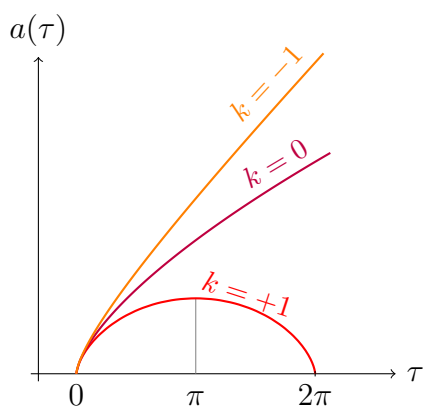
\begin{figure} 
   \centering
\begin{tikzpicture} 
  \draw[->] (-.6,0) -- (4.2,0) node[right] {$\tau$};
  \draw[->] (-.5,-.1) -- (-.5,4.2) node[above] {$a(\tau)$};
\draw[help lines] (pi/2,0) node[below,black]{$\vphantom{2}\pi$}--(pi/2,1);
\draw (pi,0) node[below] {$2\pi$};
\draw (0,0) node[below] {$0$};
\draw[domain=0:2*pi,smooth,variable=\t, red, thick] plot ({(\t-sin(\t r))/2},{(1-cos(\t r))/2});
\draw[domain=.001:pi+.2,samples=100,smooth, variable=\t, purple, thick] plot ({\t},{(exp(1/3*ln(9/4)))*exp(2/3*ln(\t))});
\draw[domain=0:pi-.2,smooth, variable=\t, orange, thick] plot ({.5*(sinh((\t))-(\t))},{.5*(cosh((\t))-1)});
\draw [shift={(pi,0)}] (0,-1pt) -- (0,1pt);
\draw [shift={(2*pi,0)}] (0,-1pt) -- (0,1pt);
\draw[red] (2.25,1.05) node[rotate=-18] {\small $k=+1$};
\draw[orange] (2.5,3.85) node[rotate=47.5] {\small $k=-1$};
\draw[purple] (2.61,2.8) node[rotate=31] {\small $k=0$};
\end{tikzpicture}
\caption{\textbf{Evoluzione del fattore di scala} dello spazio-tempo di Robertson-Walker in funzione del tempo proprio misurato dalle particelle di polvere. Si vede che per $k=0,1$ vi \`e un'espansione indefinita, invece per $k=+1$ la fase iniziale di espansione \`e seguita da una seconda fase di contrazione.}
   \label{fig:scala}
\end{figure}

Se l'universo \`e chiuso ($k=+1$) e $\rho>0$, la velocit\`a di espansione $\dot a$ pu\`o annullarsi e l'evoluzione del fattore di scala invertirsi. Questo accade se il massimo del potenziale gravitazionale assume valori sufficientemente elevati, $V_{\textsf{max}}>-1/2$ (vedi figura~\ref{fig:potcosmo}).
Per un universo chiuso di polvere e radiazione, $V_{\textsf{max}}$ si annulla asintoticamente, e questo avviene sempre.
Quando il fattore di scala raggiunge il suo valore massimo $a_{\textsf{max}}$, la velocit\`a di espansione $\dot a$ si annulla e l'universo passa da una fase di espansione a una fase di contrazione.
Negli istanti successivi, se vi \`e solamente materia ordinaria, $\ddot a<0$ e la contrazione accelera e continua fino a raggiungere inevitabilmente una singolarit\`a con $a=0$. Questa singolarit\`a futura, di stessa natura del Big Bang, si chiama \textit{Big Crunch}\index{Big Crunch}. L'universo chiuso, assumendo $\rho>0$ e $\press\geq0$, esiste dunque solamente per un tempo finito. L'evoluzione del fattore di scala per un universo di polvere ($\press=0$) \`e mostrata in figura~\ref{fig:scala} per diversi valori di $k$.

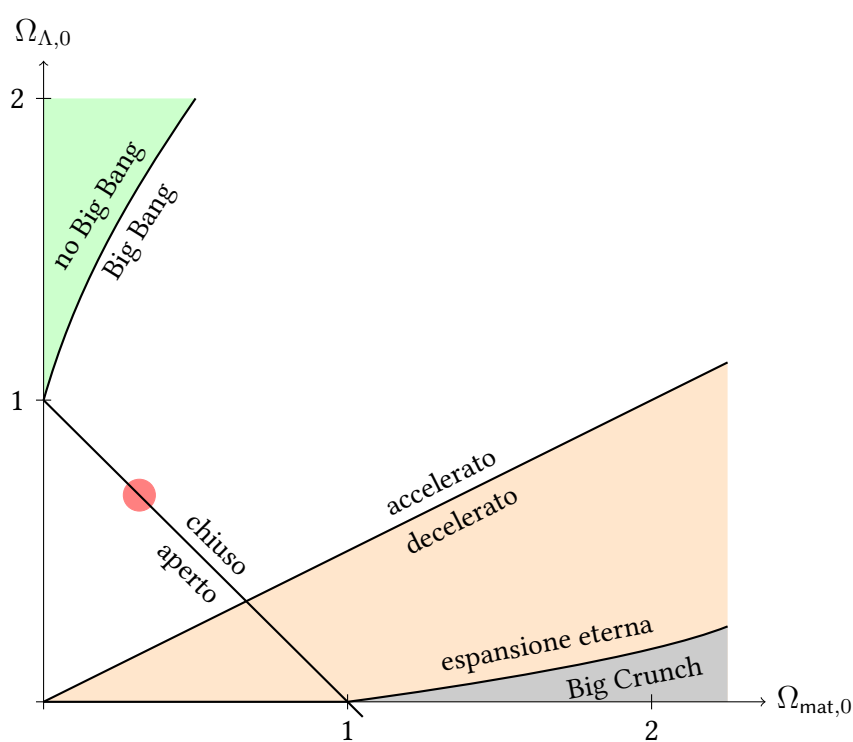
\begin{figure}[tbh] 
   \centering
   \begin{tikzpicture}

	\fill[orange!20!white] (0.0, 0.0) -- (9.0, 4.5) -- (9.0, 0.0) -- cycle;
	\fill[black!20!white] (0.0, 0.0) .. controls (1.0, 0.0) and (3.0, 0.0) .. (4.0, 0.0) .. controls (5.5, 0.2) and (8.0, 0.6) .. (9.0, 1.0) --(9.0, 0.0) -- cycle; 
	\fill[green!20!white] (0.0, 4.0) .. controls (0.4, 5.4) and (1.0, 6.53) .. (2.0, 8.0) --(0.0, 8.0) -- cycle;
	\draw[->] (-0.1,  0.0) -- ( 9.5, 0.0)
		node[right]{$\Omega_{\textsf{mat},0}$}; 
	\draw[->] ( 0.0, -0.1) -- (-0.0, 8.5)
		node[above]{$\Omega_{\Lambda,0}$}; 
	\filldraw [red!50!white] (1.26, 2.74) circle (6pt);
	\draw[thick] (0.0, 0.0) -- (9.0, 4.5)
		node[midway, above, anchor=north west, sloped]{\small decelerato}
		node[midway, below, anchor=south west, sloped]{\small accelerato};
	\draw[thick] (4.2, -0.2) -- (0.0, 4.0)
		node[midway, above, sloped]{\small chiuso}
		node[midway, below, sloped]{\small aperto};
	\draw (0.1, 4.0)--(-0.1, 4.0)
		node[left]{1};
	\draw (0.1, 8.0)--(-0.1, 8.0)
		node[left]{2};
	\draw (4.0, 0.1)--(4.0, -0.1)
		node[below]{1};
	\draw (8.0, 0.1)--(8.0, -0.1)
		node[below]{2};
	\path[thick,white] (0.0, 4.80) -- (2.0, 8.0)
		node[midway, above, sloped, black]{\small no Big Bang}
		node[midway, below, sloped, black]{\small Big Bang};
	\draw[thick] (0.0, 4.0) .. controls (0.4, 5.4) and (1.0, 6.53) .. (2.0, 8.0);
	\draw[thick] (0.0, 0.0) .. controls (1.0, 0.0) and (3.0, 0.0) .. (4.0, 0.0) .. controls (5.5, 0.2) and (8.0, 0.6) .. (9.0, 1.0)
	node[anchor = south, sloped, midway]{\small espansione eterna}
	node[anchor = north west, sloped, midway]{\small Big Crunch}
	;
\end{tikzpicture}
\caption{\textbf{Propiet\`a dell'universo in funzione del suo contenuto in materia ed energia oscura.}
Il disco rosso indica la posizione del nostro universo nel diagramma (vedi tabella~\ref{tab:density}): viviamo in un universo molto piatto, nato da un Big Bang, e la sua espansione accelerata durer\`a indefinitamente.}
   \label{fig:Omega}
\end{figure}

Come cambia questo risultato in presenza di energia oscura? Se $\Lambda<0$, e quindi $\rho_\Lambda<0$, l'equazione $V(a_{\textsf{max}})=-k/2$ ha soluzione indipendentemente dal valore di $k$ e l'universo \`e destinato a subire il Big Crunch dopo una fase di contrazione. Nel caso pi\'u interessante per la cosmologia in cui $\Lambda>0$ (e dunque $\rho_\Lambda>0$), il potenziale ha sempre un massimo $V_{\textsf{max}}$; come gi\`a accennato, 
per entrare in una fase di contrazione, l'universo deve essere chiuso e $V_{\textsf{max}}>-1/2$, altrimenti il suo fato \`e di espandersi e raffreddarsi indefinitamente.
\begin{Esercizio}
Si consideri un universo chiuso, con $\Lambda>0$ e contenente solo materia.
Mostrare che se l'universo entra in una fase di contrazione se
\eq
\Omega^{1/3}_{\Lambda,0}\Omega_{\textsf{mat},0}^{2/3}<\frac{2^{2/3}}{3}(\Omega_{\Lambda,0}+\Omega_{\textsf{mat},0}-1),
\label{V0m}\eeq
e $\Omega_{\textsf{mat},0}>2\Omega_{\Lambda,0}$, altrimenti si espande per l'eternit\`a. Similmente, mostrare che un universo chiuso con $\Lambda>0$, ma contenente solamente radiazione, entra in una fase di contrazione se e solo se
\eq
2\sqrt{\Omega_{\Lambda,0}\Omega_{\textsf{rad},0}}<\Omega_{\Lambda,0}+\Omega_{\textsf{rad},0}-1,
\label{V0r}\eeq
e  $\Omega_{\textsf{mat},0}<2\Omega_{\Lambda,0}$.
Dedurre, utilizzando i valori della tabella~\ref{tab:density} e trascurando la densit\`a di radiazione, che il nostro universo continuer\`a ad espandersi per sempre.

La presenza o meno di una singolarit\`a iniziale \`e decisa dall'altro ramo di soluzioni delle equazioni~\eqref{V0m} e~\eqref{V0r}. Mostrare che in assenza di radiazione, il Big Bang avviene se sono soddisfatte le equazioni~\eqref{V0m} e $\Omega_{\textsf{m},0}<2\Omega_{\Lambda,0}$, mentre un universo con sola radiazione ed energia oscura scaturisce da un Big Bang se sono verificate le equazioni~\eqref{V0m} e $\Omega_{\textsf{m},0}<\Omega_{\Lambda,0}$.

Riassumere i risultati ottenuti nel piano $(\Omega_{\textsf{m},0},\Omega_{\Lambda,0})$, aggiungendo la rette che separa universi chiusi e aperti, e quella che separa universi in espansione accelerata da quelli in espansione decelerata,
riproducendo cos\'\i\ il diagramma in figura~\ref{fig:Omega}.
\end{Esercizio}

Le osservazioni attuali non ci permettono di discriminare tra diversi valori di $k$, ma anche se vivessimo in un universo chiuso, con $k=+1$, il nostro universo non entrer\`a mai in una fase di contrazione. Sembra dunque che il nostro universo sia destinato a espandersi per sempre e a diventare freddo e vuoto, con un'evoluzione dettata dall'energia di vuoto.

\section{I problemi della cosmologia di Fried\-mann-Ro\-bert\-son-Wal\-ker}
\subsection{Struttura causale e orizzonti}
Abbiamo concluso, nella sezione precedente, che l'universo esiste da un tempo finito. Inoltre, sappiamo che tutti i segnali hanno una velocit\`a di propagazione limitata superiormente dalla velocit\`a della luce $c$.
Con le osservazioni abbiamo accesso solo al nostro cono luce passato, e se la radiazione che riceviamo ha avuto solo un tempo finito a disposizione per raggiungerci, \`e naturale chiedersi se possiamo vedere tutto, o solo una parte dell'universo.
In altre parole, vorremmo capire quali osservatori isotropi (altre galassie) possono aver inviato segnali che raggiungono un evento $p$ o il suo passato (e dunque possono aver influenzato gli eventi in $p$).

Per fare questo, dobbiamo studiare la \textit{struttura causale}\index{struttura causale} del nostro universo. Studieremo pi\'u avanti in dettaglio le tecniche globali che permettono questa analisi; per ora ci accontentiamo di osservare che il problema \`e semplificato dall'introduzione di una nuova coordinata temporale, il \textit{tempo conforme}\index{tempo conforme} $t$, legato
al tempo proprio $\tau$ dalla relazione
\eq
t(\tau)=\int^{\tau}\frac{d\tau'}{a(\tau')}.
\eeq
Usando questa coordinata, la metrica di Robertson-Walker~\eqref{RWmetric} diventa
\eq
ds^2=a^2(t)\left[-dt^2+
d\chi^2+\Sigma^2_k(\chi)\left(d\theta^2+\sin^2\theta\,d\varphi^2\right)\right].
\label{confRW}\eeq
Vediamo che con questa scelta di coordinate la metrica di Robertson-Walker differisce dalla metrica statica 
\eq
ds^2_{\mathsf{stat}}=-dt^2+
d\chi^2+\Sigma^2_k(\chi)\left(d\theta^2+\sin^2\theta\,d\varphi^2\right),
\label{statRW}\eeq
solo per la presenza del fattore di scala $a^2(t)$ che moltiplica tutte le componenti.
\begin{Definizione}
Si dice che due metriche $g_\munu$ e $\tilde g_\munu$ sono legate da una \textit{trasformazione conforme}\index{trasformazione conforme} se esiste una funzione $\Om(x)>0$ 
tale che 
\eq
\tilde g_\munu=\Omega^2(x)g_\munu.
\eeq
\end{Definizione}
Le trasformazioni conformi modificano localmente la scala dello spazio-tempo e in conseguenza le distanze/tempi propri tra due eventi cambiano. Tuttavia, gli angoli sono preservati e, in particolare, la relazione causale tra due eventi -- separazione di tipo tempo, spazio, o luce -- non viene modificata da una tale trasformazione.
In altre parole, i coni luce e la struttura dello spazio-tempo sono invarianti sotto trasformazioni conformi.

La struttura causale delle metriche~\eqref{confRW} e~\eqref{statRW} \`e dunque la stessa, e possiamo dunque analizzarla nella metrica statica dove il problema risulta pi\'u semplice.
Senza perdita di generalit\`a studiamo le geodetiche radiali nulle; l'annullamento dell'intervallo~\eqref{statRW} implica che queste verificano $\dd t=\pm\dd\chi$, e in conseguenza se un fotone emesso in $(t_{\mathsf{e}},\chi_{\mathsf{e}})$ viene ricevuto in $(t_{\mathsf{r}},\chi_{\mathsf{r}})$, questi eventi verificano
\eq
t_{\mathsf{r}}-t_{\mathsf{e}}=\left|\chi_{\mathsf{r}}-\chi_{\mathsf{e}}\right|.
\eeq
Indicando con $t_{\mathsf{i}}$ il tempo conforme al momento del Big Bang, l'osservatore pu\`o dunque ricevere segnali da una distanza comovente massima
\eq
\left|\chi_{\mathsf{r}}-\chi_{\mathsf{e}}\right|\leq \Delta\chi_{\mathsf{max}}= t_{\mathsf{r}}-t_{\mathsf{i}}
\quad\textsf{(Et\`a dell'universo al tempo di ricezione)}.
\eeq
Segnali emessi da una distanza comovente superiore a $\Delta\chi_{\mathsf{max}}$ non hanno avuto tempo di raggiungere l'evento $p$. Esistono dunque regioni dell'universo inaccessibili all'osservazione.
\begin{Definizione}
La frontiera tra le linee di mondo che possono essere viste in un evento $p$ e quelle che non possono si chiama \textit{orizzonte delle particelle}
\index{orizzonte delle particelle}
\index{orizzonte degli oggetti|see{orizzonte delle particelle}}
(o degli oggetti) in $p$.
\end{Definizione}
Visto che il fattore di scala si annulla sulla singolarit\`a iniziale, nei primissimi istanti dopo il Big Bang tutti i punti erano estremamente vicini, e si potrebbe pensare che i segnali potevano propagare facilmente da un evento all'altro. Ma questo non \`e vero se il fattore di scala cresce velocemente. Infatti, l'osservatore in $p$ riceve segnali da osservatori arbitrariamente lontani solamente se $t_{\mathsf{i}}\to-\infty$, ovvero se l'integrale
\eq
t-t_{\mathsf{i}}=\int_{\tau_{\mathsf{i}}}^{\tau}\frac{d\tau}{a(\tau)}\quad\text{diverge.}
\label{hint}\eeq
Questo avviene se esiste una costante $\alpha$ tale che $a(\tau)\leq\alpha(\tau-\tau_{\mathsf{i}})$ per $\tau\to\tau_{\mathsf{i}}$, ovvero se $a(\tau)$ tende a zero sufficientemente velocemente quando ci avviciniamo al Big Bang. In questo caso, non vi \`e orizzonte delle particelle: le linee di mondo di tutti gli osservatori intersecano il cono luce passato di un determinato evento $p$, e lo spazio-tempo di Robertson-Walker \`e conformemente legato all'intero spazio-tempo statico \eqref{statRW}.

\begin{figure}
\centering
\begin{tikzpicture}
\fill[gray] (0,2.5) circle(2pt) node[black,anchor=south west]{$p$};
\fill[blue!5!white] (-2.5,0)--(0,2.5)--(2.5,0)--cycle;
\draw[blue] (-2.5,0)--(0,2.5)--(2.5,0);
\draw[very thick, decorate, decoration={random steps,segment length=3pt,amplitude=1pt}] (-4,0)--(4,0) node[right]{$t_{\mathsf{i}}$};
\draw[help lines] (-4,2.5)--(4,2.5) node[right, black]{$t_{0}$};
\draw[->, help lines] (-2.5,0) node [below] {\small $\chi=-|t_0-t_{\mathsf{i}}|$} -- (-2.5,3);
\draw[->, help lines] (0,0) node [below] {\small $\chi=0$} -- (0,3);
\draw[->, help lines] (2.5,0) node [below] {\small $\chi=|t_0-t_{\mathsf{i}}|$} -- (2.5,3);
\end{tikzpicture}
\caption{\textbf{Orizzonte delle particelle.}}
\label{fig:orizzonte}\end{figure}
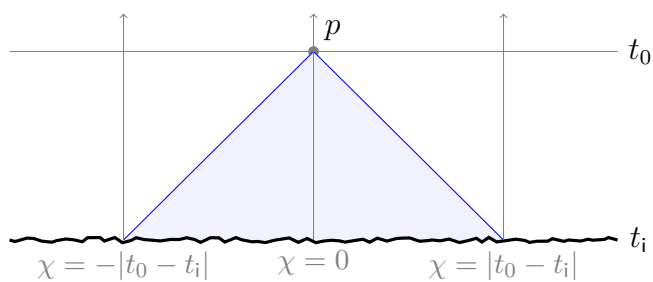

Se invece l'integrale~\eqref{hint} converge, lo spazio-tempo di Robertson-Walker copre solo il semispazio $t>t_{\mathsf{i}}$ dello spazio-tempo statico \eqref{statRW}, e vi \`e un orizzonte (vedi figura~\ref{fig:orizzonte}).
Nelle teorie cosmologiche convenzionali \`e sempre presente un'orizzonte delle particelle. Infatti, indipendentemente dal contributo di materia ed energia oscura al tempo presente, la densit\`a di energia era dominata dalla radiazione nel passato, e dunque (con $a(\tau)\propto\sqrt{\tau}$, vedi tabella~\ref{tab:a}) l'integrale~\eqref{hint} converge.

La distanza propria dall'orizzonte \`e dunque data da
\eq
d_{\mathsf{max}}(\tau)=a(\tau)\int^{\chi_{\mathsf{max}}(\tau)}_0
d\chi
=a(\tau)\int_{\tau_{\mathsf i}}^\tau\frac{d\tau'}{a(\tau')}.
\eeq

\textbf{I modelli di Robertson-Walker con $w>-1/3$ hanno un orizzonte.}
Cambiando variabile d'integrazione, la distanza dall'orizzonte si pu\`o riscrivere
\eq
d_{\mathsf{max}}=a_0\int_0^{a_0}\frac{da}{a\dot a},
\label{dH}\eeq
dove abbiamo usato $a(\tau_{\mathsf i})=0$. Per semplicit\`a, consideriamo un universo piatto, $k=0$. Supponiamo inoltre che l'evoluzione dell'universo sia dominata da una singola componente con equazione di stato $P=w\rho$ nei primissimi istanti dopo il Big Bang. La densit\`a di energia \`e dunque data da~\eqref{rho(a)} in quel periodo e la prima equazione di Friedmann~\eqref{F1} implica che
\eq
a\dot a\propto a^{\frac{1-3w}2}.
\eeq
Perci\`o nell'espressione~\eqref{dH} integriamo la funzione $a^{(3w-1)/2}$ vicino all'origine. Questa operazione risulta in una divergenza per $w\leq-1/3$, ma altrimenti rimane finita. Conseguentemente, quando $w>-1/3$, e in particolare per un universo di polvere ($w=0$) e di radiazione ($w=1/3$), l'orizzonte si trova a una distanza finita.
Questa conclusione non viene modificata dalla presenza di una costante cosmologica, in quanto l'evoluzione primordiale dell'universo \`e dettata -- come abbiamo visto in precedenza -- dalla componente del fluido cosmologico con $w$ pi\`u grande, ed \`e il comportamento vicino alla singolarit\`a iniziale del fattore di scala che determina la presenza di orizzonti.

Se la geometria dell'universo \`e sferica ($k=+1$) o iperbolica $(k=-1)$ abbiamo l'ulteriore termine di curvatura $k/a^2$ nella prima equazione di Friedmann~\eqref{F1}. Tuttavia, per $a\to0$, questo termine diventa trascurabile rispetto alla densit\`a di energia se $w>-1/3$, e dunque il comportamento \`e ancora una volta simile al caso piatto $k=0$: vi \`e sempre un orizzonte degli oggetti. Per\`o, se l'universo chiuso \`e compatto, \`e anche possibile che l'orizzonte cresca fino a raggiungere un raggio pari una semi-circonferenza della 3-sfera spaziale ($\Delta\chi=\pi$ in coordinate comoventi). Questo avviene quando un segnale luminoso emesso al momento del Big Bang raggiunge il punto antipodale sulla 3-sfera. Da quel momento in poi, un qualunque osservatore pu\`o ricevere segnali da tutti gli osservatori isotropi, e l'orizzonte degli oggetti sparisce.
\begin{Esercizio}
Si consideri un universo chiuso ($k=+1$) con sezioni spaziali sferiche.
\begin{enumerate}
\item Si consideri un una geodetica radiale di tipo luce emessa al momento $t_{\mathsf{i}}$ del Big Bang. Si calcoli la distanza comovente $\Delta\chi(t)$ percorsa da questa geodetica al momento $t$.
\item Si mostri che per un universo di polvere un raggio di luce emesso al Big Bang ha fatto esattamente un giro della 3-sfera al momento del Big Crunch. Analogamente, si mostri che nel caso di universo di radiazione il raggio di luce ha percorso esattamente la met\`a di un cerchio massimo quando raggiunge la singolarit\`a finale.
\item Dedurre che per un universo di polvere l'orizzonte sparisce nel momento in cui l'universo raggiunge il suo raggio massimo ed entra nella fase di contrazione, mentre per l'universo di polvere esso sparisce al momento del Big Crunch.
\item Mostrare che le sezioni $(\theta,\phi)$ costante sono conformi a una regione finita dello spazio-tempo di Minkowski in due dimensioni. Tracciare il diagramma corrispondente indicando il Big Bang, Big Crunch, la propagazione delle geodetiche nulle e illustrare il risultato ottenuto al punto precedente.
\end{enumerate}
\end{Esercizio}

Attualmente la densit\`a di energia nel nostro universo \`e originata da una componente di energia oscura $\rho_\La$, una componente di materia $\rhom$ e una componente di radiazione $\rhor$, e la sua evoluzione \`e dominata da $\rho_\La$. Tuttavia, abbiamo visto che durante l'evoluzione cosmologica $\rho_\La$ rimane costante, mentre $\rhom$ e $\rhor$ decrescono come $a^{-3}$ e $a^{-4}$ rispettivamente (vedi figura~\ref{fig:comp}). Inevitabilmente, in un modello di questo tipo, ai primordi l'evoluzione dell'universo era guidata dalla radiazione, e dunque, per quanto appena visto, vi \`e un orizzonte. Come stima della distanza dell'orizzonte possiamo prendere la distanza di Hubble\index{Hubble!distanza di} $d_{\mathsf{H}}=cH_0^{-1}$ che corrisponde alla distanza propria degli osservatori isotropi pi\'u lontani che hanno potuto inviarci segnali, approssimati con la distanza percorsa dalla luce durante il tempo di Hubble\index{Hubble!tempo di} (che corrisponde all'incirca all'et\`a dell'universo). Questa approssimazione \`e ragionevole per un universo piatto formato da materia ordinaria ($\rho\geq0$).
\begin{Esercizio} Mostrare che, per un Universo piatto contenente un'unica componente di energia con equazione di stato $P=w\rho$ e $w\geq-1/3$,
\eq
d_{\mathsf{max}}=\frac2{1+3w}H_0^{-1},
\eeq
giustificando cos\'\i\ la definizione della distanza di Hubble.
\end{Esercizio}

\subsection{Il problema dell'orizzonte}
L'alto grado di uniformit\`a della radiazione cosmica di fondo che viene osservata \`e difficile da riconciliare con la presenza di un orizzonte delle particelle. Infatti, l'isotropia della CMB significa che al momento del disaccoppiamento -- quando i fotoni hanno interagito per l'ultima volta con il plasma che permeava l'universo prima di intraprendere il loro lungo viaggio fino ad essere catturati dai nostri sensori -- il plasma era all'equilibrio termico, con fluttuazioni in temperatura molto piccole da un punto all'ultro, dell'ordine $\Delta T/T\sim10^{-5}$. Per\`o, per raggiungere l'equilibrio termico, le varie regioni di un sistema devono aver potuto interagire per un tempo sufficientemente lungo affinch\'e esso possa dissipare le inomogeneit\`a e rilassare all'equilibrio termico.

Visto che l'universo non \`e esistito per un tempo infinito, ma al contrario \`e emerso dal Big Bang al tempo $t_{\mathsf i}$, dobbiamo chiederci se la radiazione di microonde che misuriamo in direzioni opposte della volta celeste provenga da regioni che hanno avuto tempo di mettersi in contatto causale, e quindi termalizzare, oppure no.

Per rispondere a questa domanda, possiamo stimare la dimensione delle regioni che hanno avuto tempo per raggiungere l'equilibrio termico al momento del disaccoppiamento, e questo lo faremo approssimandola con la distanza che la luce emessa al momento del Big Bang ha potuto percorrere prima di quel tempo. Confronteremo poi questa distanza con le dimensioni dell'orizzonte di Hubble, che corrisponde approssimativamente alla distanza percorsa dai fotoni della CMB prima di raggiungerci.

Nel caso di un universo composto da un fluido con parametro $w$, otteniamo la distanza comovente percorsa da un segnale luminoso emesso quando il fattore di scala assumeva il valore $a_1$ e il momento con fattore di scala $a_2$ dalla relazione~\eqref{dH} che, effettuato l'integrale, fornisce
\eq
\Delta\chi=\int^{a_2}_{a_1}\frac{da}{a\dot a}=\frac{2\chi_{\textsf H}}{3w+1}\left[\lp\frac{a_2}{a_0}\rp^{\frac{3w+1}2}-\lp\frac{a_1}{a_0}\rp^{\frac{3w+1}2}\right],
\label{chi}\eeq
dove abbiamo definito il raggio di Hubble comovente
\eq
\chi_{\textsf H}=\frac{1}{H_0a_0}.
\label{HubbleComoving}
\eeq
Quindi $\Delta\chi=\chi_{\textsf H}(a_2-a_1)$ per un universo dominato da radiazione, e $\Delta\chi=2\chi_{\textsf H}(\sqrt{a_2}-\sqrt{a_1})$ per un universo dominato da materia. Ricordiamo che la materia \`e diventata dominante per $z_{\mathsf{mat}}=8800$ e che il disaccoppiamento tra radiazione e materia \`e avvenuto per $z_{\mathsf{dec}}=1100$. Approssimando l'intera evoluzione del fattore di scala con un'evoluzione dominata dalla materia non relativistica ($w=0$) e utilizzando la relazione tra fattore di scala e redshift \eqref{zdef}, troviamo che la dimensione dell'orizzonte al momento del disaccoppiamento \`e
\eq
\Delta\chi_{\mathsf{dec}}=\frac2{\sqrt{1+z_{\mathsf{dec}}}}.
\eeq
Dobbiamo confrontare questa quatit\`a con il raggio di Hubble al disaccoppiamento, ovvero la distanza comovente che ha potuto percorrere un segnale al momento del disaccoppiamento. Dalla~\eqref{chi} troviamo
\eq
\Delta\chi_{\textsf{H}}=2\chi_{\textsf H}\lp1-\frac1{\sqrt{1+z_{\textsf{dec}}}}\rp,
\eeq
e dunque
\eq
\frac{\Delta\chi_{\textsf{dec}}}{\Delta\chi_{\textsf{H}}}\approx0.03\,\textsf{rad}\approx1.7^{\,\circ}.
\label{dimensioniHubbleDec}
\eeq
Dunque, al momento della ricombinazione, solo piccole regioni -- che oggi viste dalla Terra sottendono un angolo di $1.7^{\,\circ}$ -- hanno avuto tempo di mettersi in contatto termico.
\begin{figure}
\centering
\begin{tikzpicture}
\fill[gray] (0,3) circle(2pt) node[black,anchor=south west]{$p$};
\fill[blue!5!white] (-3,0)--(0,3)--(3,0)--cycle;
\fill[red!7!white, path fading=fade up] (-2.25,1)--(-2,.75)--(-.5,.75)--(-.25,1)--cycle;
\fill[red!7!white] (-2,.75)--(-1.25,0)--(-.5,.75)--cycle;
\draw[red, thin] (-2,.75)--(-1.25,0)--(-.5,.75);
\draw[red, path fading=fade up, thin] (-2,.75)--(-2.25,1);
\draw[red, path fading=fade up, thin] (-.5,.75)--(-.25,1);
\fill[red!7!white, path fading=fade up, shift={(2.2,0)}] (-2.25,1)--(-2,.75)--(-.5,.75)--(-.25,1)--cycle;
\fill[red!7!white, shift={(2.2,0)}] (-2,.75)--(-1.25,0)--(-.5,.75)--cycle;
\draw[red, thin, shift={(2.2,0)}] (-2,.75)--(-1.25,0)--(-.5,.75);
\draw[red, path fading=fade up, thin, shift={(2.2,0)}] (-2,.75)--(-2.25,1);
\draw[red, path fading=fade up, thin, shift={(2.2,0)}] (-.5,.75)--(-.25,1);
\draw[help lines,name path = tdec] (-4,.75) node[black, anchor=east]{\hphantom{$t_{\textsf{dec}}$}} -- (4,.75) node[black, anchor=west]{$t_{\textsf{dec}}$};
\draw[blue,name path=hleft] (-3,0)--(0,3);
\draw[blue,name path=hright] (0,3)--(3,0);
\draw[very thick, decorate, decoration={random steps,segment length=3pt,amplitude=1pt}] (-4,0)--(4,0) node[right]{$t_{\mathsf{i}}$};
\draw[help lines] (-4,3)--(4,3) node[right, black]{$t_{0}$};
\draw[help lines,->] (-4,1.5)--(-3.2,1.5) node[below,black] {\footnotesize $\chi$};
\draw[help lines,->] (-4,1.5)--(-4,2.3) node[left,black] {\footnotesize $t$};
\fill[gray] (-4,1.5) circle(1.5pt);
%

\draw [help lines, dashed, name intersections={of=tdec and hright, by={a}}, name intersections={of=tdec and hleft, by={b}}]
    (a |- 0, -0.4) -- (a |- 0, 0.9);
\draw [help lines, dashed, name intersections={of=tdec and hleft, by={b}}]
    (b |- 0, -0.4) -- (b |- 0, 0.9);
\draw[help lines,<->] (a |- 0, -0.3) -- (b |- 0, -0.3) node[midway, below, black] {$\Delta\chi_{\textsf H}$};

\draw[help lines] (0.2, 0.6)--(0.2, 1);
\draw[help lines] (1.7, 0.6)--(1.7, 1);
\draw[help lines,<->] (0.2, 0.9) -- (1.7, 0.9) node[midway, above, black] {$\Delta\chi_{\textsf{dec}}$};
\end{tikzpicture}
\caption{\textbf{Problema dell'orizzonte.} Al momento $t_{\textsf{dec}}$ del disaccoppiamento, le regioni all'equilibrio termico hanno dovuto prima essere in contatto causale (in rosso nel diagramma). Ma queste regioni sottendono solo una frazione della volta celeste che osserviamo oggi. }
\label{fig:problemaorizzonte}\end{figure}
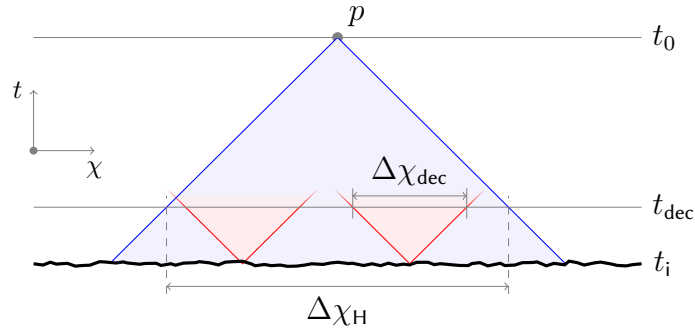
La temperatura della radiazione di fondo cosmico che osserviamo \`e invece isotropa, e implica che una regione di dimensioni $\Delta\chi_{\mathsf{h}}\gg\Delta\chi_{\mathsf{dec}}$ doveva essere all'equilibrio termico al momento del disaccoppiamento. Ma questa regione non ha avuto tempo di mettersi tutta in contatto causale -- e dunque di raggiungere l'equilibrio termico -- prima dell'epoca del disaccoppiamento.
L'impossibilit\`a di spiegare l'isotropia e omogeneit\`a della radiazione cosmica di fondo nell'ambito del modello di Friedmann-Robertson-Walker \`e insoddisfacente, e costituisce il problema dell'orizzonte. Questo richiede di imporre in modo poco naturale condizioni iniziali estremamente omogenee e isotrope per l'evoluzione dell'universo. Una possibilit\`a alternativa \`e che l'evoluzione dell'universo durante i primi istanti dopo il Big Bang era molto diversa da quella prevista dal modello di Friedmann-Robertson-Walker. Vedremo in che una fase iniziale di espansione accelerata permette di superare queste difficolt\`a, in quanto permette di mettere in contatto causale e di termalizzare a tutto l'universo osservabile entro il momento del disaccoppiamente, e inoltre sopprime tutte le disomogeneit\`a presenti inizialmente.

\subsection{Il problema della piattezza}
\label{sec:piattezza}

Il parametro di densit\`a stimato dell'universo attuale \`e molto vicino a $\Omega_{\mathsf{tot},0}=1$, ovvero alla densit\`a critica corrispondente a un universo piatto,
\beq
\Omega_{\mathrm{tot},0}=1\pm 0.007\,.
\label{Omega}\eeq
Nel modello di Friedmann-Robertson-Walker, questo valore del parametro di densit\`a \`e poco naturale, in quanto l'evoluzione cosmologica allontana il parametro densit\`a dal valore critico, e richiede un aggiustamento fine\index{aggiustamento fine} (o \textit{fine tuning}\index{fine tuning}) delle condizioni iniziali per l'universo primordiale, fatto non soddisfacente da un punto di vista teorico. Per vedere come si presenta questo problema -- e capire se \`e possibile ovviarlo -- studiamo l'evoluzione temporale di $\Omega$. Riscriviamo la prima equazione di Friedmann~\eqref{F1} nella forma
\beq
\Omega=1+\frac k{a^2H^2}\,,
\eeq
e attraverso una riparametrizzazione della coordinata temporale, studiamo la variazione di $\Omega$ al variare del fattore di scala,
\beq
\frac{d\Omega}{da}=-{2k}{a^2H^2}\lp\frac1a+\frac1H\frac{dH}{da}\rp.
\eeq
D'altra parte, utilizzando la seconda equazione di Friedmann~\eqref{F2},
\beq
\frac{dH}{da}=\frac1{\dot a}\frac{dH}{d\tau}=-\frac1{aH}\left[\frac{4\pi\newt}3\lp\rho+3P\rp+H^2\right],
\eeq
e dunque, al variare del fattore di scala dell'universo, l'evoluzione del parametro di densit\`a \`e dettata dall'equazione\footnote
{
Questa equazione differenziale, della forma $y'=y(1-y)$, si dice \textit{equazione logistica}.
}
\beq
\frac{d\Omega}{d \log a}=\Omega(\Omega-1)\lp1+\frac{3P}\rho\rp.
\label{eqOmega}\eeq

\begin{figure}
\centering
\begin{tikzpicture} \begin{axis}[axis equal image=true, axis x line=bottom,
    axis y line=left, ymin=0, ymax=2.48, domain=0:4, samples=20, smooth, ultra thin,
	xtick={1,2,3}, ytick={0,1}, xlabel=$\ln{a}$,ylabel=$\Omega$,
	every axis x label/.style={at={(current axis.right of origin)},anchor=north},
    every axis y label/.style={at={(current axis.above origin)},anchor=north east},
    yticklabel=\ , xticklabel=\ , 
    extra description/.code={\node[left] at (axis cs:0,0) {$0$};
                             \node[left] at (axis cs:0,1) {$1$};}
]
\addplot[blue!80!black, postaction={decorate}, decoration={markings,
         mark=at position 0.1 with {\arrow{stealth}}}]{1/(1+exp((1+3*1/3)*x+2))};
\addplot[blue!80!black, postaction={decorate}, decoration={markings,
         mark=at position 0.1 with {\arrow{stealth}},
         mark=at position 0.3 with {\arrow{stealth}}}]{1/(1+exp((1+3*1/3)*x))};
\addplot[blue!80!black, postaction={decorate}, decoration={markings,
         mark=at position 0.1 with {\arrow{stealth}},
         mark=at position 0.3 with {\arrow{stealth}},
         mark=at position 0.5 with {\arrow{stealth}}}]{1/(1+exp((1+3*1/3)*x-2))};
\addplot[blue!80!black, domain=0:3.90,postaction={decorate}, decoration={markings,
         mark=at position 0.31 with {\arrow{stealth}},
         mark=at position 0.51 with {\arrow{stealth}},
         mark=at position 0.71 with {\arrow{stealth}}}]{1/(1+exp((1+3*1/3)*x-4))};
\addplot[blue!80!black, postaction={decorate}, decoration={markings,
         mark=at position 0.5 with {\arrow{stealth}},
         mark=at position 0.7 with {\arrow{stealth}},
         mark=at position 0.9 with {\arrow{stealth}}}]{1/(1+exp((1+3*1/3)*x-6))};
\addplot[blue!80!black, postaction={decorate}, decoration={markings,
         mark=at position 0.7 with {\arrow{stealth}},
         mark=at position 0.9 with {\arrow{stealth}}}]{1/(1+exp((1+3*1/3)*x-8))};
\addplot[blue!80!black, postaction={decorate}, decoration={markings,
         mark=at position 0.9 with {\arrow{stealth}}}]{1/(1+exp((1+3*1/3)*x-10))};
\addplot[blue!80!black, postaction={decorate}, decoration={markings,
         mark=at position 0.95 with {\arrow{stealth}}}]{1/(1-exp((1+3*1/3)*x-10))};
\addplot[blue!80!black, domain=0:4, postaction={decorate}, decoration={markings,
         mark=at position 0.65 with {\arrow{stealth}},
         mark=at position 0.8 with {\arrow{stealth}}}]{1/(1-exp((1+3*1/3)*x-8.5))};
\addplot[blue!80!black, domain=0:3.3, postaction={decorate}, decoration={markings,
         mark=at position 0.6 with {\arrow{stealth}},
         mark=at position 0.8 with {\arrow{stealth}}}]{1/(1-exp((1+3*1/3)*x-7))};
\addplot[blue!80!black, domain=0:2.55, postaction={decorate}, decoration={markings,
         mark=at position 0.58 with {\arrow{stealth}},
         mark=at position 0.8 with {\arrow{stealth}}}]{1/(1-exp((1+3*1/3)*x-5.5))};
\addplot[blue!80!black, domain=0:1.8, postaction={decorate}, decoration={markings,
         mark=at position 0.55 with {\arrow{stealth}},
         mark=at position 0.78 with {\arrow{stealth}}}]{1/(1-exp((1+3*1/3)*x-4))};
\addplot[blue!80!black, domain=0:1.05, postaction={decorate}, decoration={markings,
         mark=at position 0.5 with {\arrow{stealth}},
         mark=at position 0.7 with {\arrow{stealth}}}]{1/(1-exp((1+3*1/3)*x-2.5))};
\addplot[blue!80!black, domain=0:0.3, postaction={decorate}, decoration={markings,
         mark=at position 0.4 with {\arrow{stealth}}}]{1/(1-exp((1+3*1/3)*x-1))};
\addplot[red!80!black, semithick, postaction={decorate}, decoration={markings,
         mark=at position 0.1 with {\arrow{stealth}},
         mark=at position 0.3 with {\arrow{stealth}},
         mark=at position 0.5 with {\arrow{stealth}},
         mark=at position 0.7 with {\arrow{stealth}},
         mark=at position 0.9 with {\arrow{stealth}}}]{1};
\end{axis}
\end{tikzpicture}
\caption{\textbf{Repulsore per $P>-\rho/3$.} In presenza di materia ordinaria e radiazione, la soluzione critica $\Omega=1$ \`e un \emph{punto fisso instabile} e l'evoluzione cosmologica allontana rapidamente il parametro densit\`a dal suo valore critico.}
\label{fig:repulsore}
\end{figure}
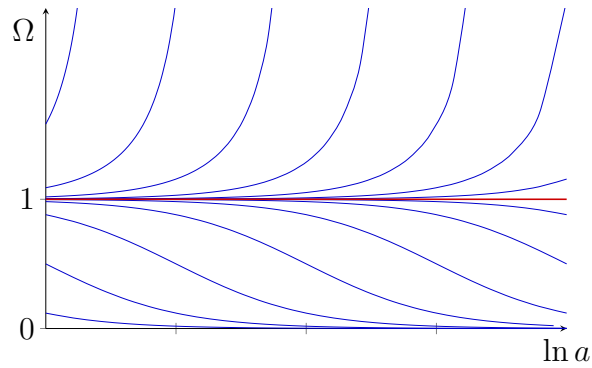
Vediamo dunque che la soluzione $\Omega=1$ \`e un punto fisso.
Per $P>-\rho/3$, e in particolare in presenza di materia ordinaria e radiazione, $\Omega$ \`e una funzione crescente di $a$ per $\Omega>1$, ed \`e una funzione decrescente di $a$ per $\Omega<1$: l'evoluzione cosmologica allontana rapidamente il parametro di densit\`a dalla soluzione $\Omega=1$. La soluzione $\Omega=1$ \`e un \textit{repulsore\index{repulsore}} dell'equazione differenziale (vedi figura~\ref{fig:repulsore}). Infatti, prendendo per esempio un'unica componente del fluido cosmologico, con equazione di stato $P=w\rho$, la soluzione
\beq
\frac{|\Omega(a)-1|}{\Omega(a)}=\frac{|\Omega_0-1|}{\Omega_0}\lp\frac{a}{a_0}\rp^{1+3w}
\eeq
dell'equazione~\eqref{eqOmega} richiede che per avere il valore attuale di $\Omega$ molto vicino al valore critico~\eqref{Omega} il parametro di densit\`a doveva avvinarsi a 1 con precisione elevatissima nel passato. Considerando per semplicit\`a un universo contenente sola radiazione,\footnote{Le correzioni dovute alla presenza di materia e energia di vuoto sono comunque trascurabili.} abbiamo $w=1/3$ e $a(\tau)\propto\sqrt{\tau}$, e possiamo approssimare
\eq
|\Omega(\tau)-1|\approx|\Omega_0-1|\frac{\tau}{\tau_0}.
\eeq
Prendendo $|\Omega_0-1|\approx10^{-3}$ e $\tau_0\approx10^{17}\,\textsf{s}$, troviamo che all'inizio della nucleosintesi, quando l'et\`a dell'universo era $\tau_{\textsf{n}}\approx1\,\textsf{s}$, il parametro densit\`a poteva discostarsi da valore critico~$1$ al pi\'u per una parte su $10^{20}$! Andando indietro fino all'epoca di Planck ($\tau_{\textsf{P}}\approx10^{-44}\,\textsf{s}$ dopo il Big Bang), troviamo $|\Omega-1|<10^{-64}$. Il valore osservato~\eqref{Omega} del parametro di densit\`a richiede dunque condizioni iniziali al momento del Big Bang altamente omogenee, e con un aggiustamento fine della densit\`a di energia poco naturale; occorre chiedersi per quale motivo, dopo un'espansione dell'universo durata miliardi di anni, il parametro di densit\`a \`e tutt'ora cos\'\i{} vicino al valore critico.

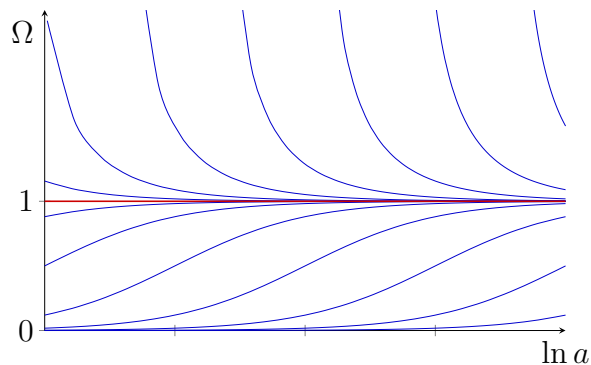
\begin{figure}
\centering
\begin{tikzpicture} \begin{axis}[axis equal image=true, axis x line=bottom,
    axis y line=left, ymin=0, ymax=2.48, domain=0:4, samples=20, smooth, ultra thin,
	xtick={1,2,3}, ytick={0,1}, xlabel=$\ln{a}$,ylabel=$\Omega$,
	every axis x label/.style={at={(current axis.right of origin)},anchor=north},
    every axis y label/.style={at={(current axis.above origin)},anchor=north east},
    yticklabel=\ , xticklabel=\ , 
    extra description/.code={\node[left] at (axis cs:0,0) {$0$};
                             \node[left] at (axis cs:0,1) {$1$};}
]
\addplot[blue!80!black, postaction={decorate}, decoration={markings,
         mark=at position 0.1 with {\arrow{stealth}}}]{1/(1+exp(-2*x-2))};
\addplot[blue!80!black, postaction={decorate}, decoration={markings,
         mark=at position 0.1 with {\arrow{stealth}},
         mark=at position 0.3 with {\arrow{stealth}}}]{1/(1+exp(-2*x))};
\addplot[blue!80!black, postaction={decorate}, decoration={markings,
         mark=at position 0.1 with {\arrow{stealth}},
         mark=at position 0.3 with {\arrow{stealth}},
         mark=at position 0.5 with {\arrow{stealth}}}]{1/(1+exp(-2*x+2))};
\addplot[blue!80!black, postaction={decorate}, decoration={markings,
         mark=at position 0.3 with {\arrow{stealth}},
         mark=at position 0.5 with {\arrow{stealth}},
         mark=at position 0.7 with {\arrow{stealth}}}]{1/(1+exp(-2*x+4))};
\addplot[blue!80!black, postaction={decorate}, decoration={markings,
         mark=at position 0.5 with {\arrow{stealth}},
         mark=at position 0.7 with {\arrow{stealth}},
         mark=at position 0.9 with {\arrow{stealth}}}]{1/(1+exp(-2*x+6))};
\addplot[blue!80!black, postaction={decorate}, decoration={markings,
         mark=at position 0.7 with {\arrow{stealth}},
         mark=at position 0.9 with {\arrow{stealth}}}]{1/(1+exp(-2*x+8))};
\addplot[blue!80!black, postaction={decorate}, decoration={markings,
         mark=at position 0.9 with {\arrow{stealth}}}]{1/(1+exp(-2*x+10))};
\addplot[blue!80!black, domain=0:4, postaction={decorate}, decoration={markings,
         mark=at position 0.1 with {\arrow{stealth}}}]{1/(1-exp(-2*x-2))};
\addplot[blue!80!black, domain=0.02:4, postaction={decorate}, decoration={markings,
         mark=at position 0.1 with {\arrow{stealth}},
         mark=at position 0.28 with {\arrow{stealth}}}]{1/(1-exp(-2*x-.5))};
\addplot[blue!80!black, domain=.7:4, postaction={decorate}, decoration={markings,
         mark=at position 0.2 with {\arrow{stealth}},
         mark=at position 0.33 with {\arrow{stealth}}}]{1/(1-exp(-2*x+1))};
\addplot[blue!80!black, domain=1.45:4, postaction={decorate}, decoration={markings,
         mark=at position 0.21 with {\arrow{stealth}},
         mark=at position 0.35 with {\arrow{stealth}},
         mark=at position 0.5 with {\arrow{stealth}}}]{1/(1-exp(-2*x+2.5))};
\addplot[blue!80!black, domain=2.2:4, postaction={decorate}, decoration={markings,
         mark=at position 0.24 with {\arrow{stealth}},
         mark=at position 0.5 with {\arrow{stealth}}}]{1/(1-exp(-2*x+4))};
\addplot[blue!80!black, domain=2.95:4, postaction={decorate}, decoration={markings,
         mark=at position 0.3 with {\arrow{stealth}},
         mark=at position 0.55 with {\arrow{stealth}}}]{1/(1-exp(-2*x+5.5))};
\addplot[blue!80!black, domain=3.7:4, postaction={decorate}, decoration={markings,
         mark=at position 0.5 with {\arrow{stealth}}}]{1/(1-exp(-2*x+7))};
\addplot[red!80!black, semithick, postaction={decorate}, decoration={markings,
         mark=at position 0.1 with {\arrow{stealth}},
         mark=at position 0.3 with {\arrow{stealth}},
         mark=at position 0.5 with {\arrow{stealth}},
         mark=at position 0.7 with {\arrow{stealth}},
         mark=at position 0.9 with {\arrow{stealth}}}]{1};
\end{axis}
\end{tikzpicture}
\caption{\textbf{Attrattore per $P<-\rho/3$.} In questo caso la soluzione critica $\Omega=1$ diventa un \emph{punto fisso stabile} e l'evoluzione cosmologica avvicina rapidamente il parametro densit\`a al suo valore critico. Questo accade ad esempio se l'energia di vuoto ($w=-1$) guida l'espansione dell'universo.}
\label{fig:attrattore}
\end{figure}

Al contrario, se $P>-\rho/3$, il punto fisso $\Omega=1$ diventa un \textit{attrattore}\index{attrattore}, e l'evoluzione cosmologica guida rapidamente il parametro di densit\`a verso il valore critico $\Omega=1$ corrispondente a un universo piatto (vedi figura~\ref{fig:attrattore}). Dall'equazione di Friedmann~\eqref{F2}, si vede che questo avviene proprio quando l'espansione dell'universo \`e accelerata.
Ne deduciamo che -- come nel caso del problema dell'orizzonte -- se nei primi istanti l'universo attraversa una fase accelerata $\ddot a>0$, il parametro di densit\`a viene portato in modo naturale vicino al valore critico. Se questa fase dura sufficientemente, la successiva espansione decelerata dell'universo, guidata da radiazione, poi materia, parte da un $\Omega$ sufficientemente vicino ad 1 per spiegare il valore odierno~\eqref{Omega} senza la necessit\`a dell'aggiustamento fine. Questa fase accelerata iniziale, durante la quale le dimensioni dell'universo aumentano di un fattore molto elevato, viene chiamata \emph{inflazione cosmica}\index{inflazione}, ed \`e l'argomento della prossima sezione.

\section{L'inflazione cosmica}
In un universo di materia ordinaria ($\rho\geq0$), il raggio comovente dell'orizzonte di Hubble~\eqref{HubbleComoving} cresce con il passare del tempo, in quanto regioni sempre pi\'u lontane dell'universo avranno avuto modo di mettersi in contatto causale con noi. Infatti, segue dalla seconda equazione di Friedmann~\eqref{F2} che con un tale contenuto di materia l'espansione \`e decelerata ($\ddot a<0$), e dunque
\eq
\frac{d\chi_{\textsf{H}}}{d\tau}=-\frac1a\lp\frac{\dot H}{H^2}+1\rp
=-\frac{\ddot a}{\dot a^2}>0.
\label{dchidtau}
\eeq
Questa crescita porta il raggio comovente di Hubble ad essere oggi molto grande rispetto alle dimensioni dell'orizzonte al momento del disaccoppiamento (vedi l'equazione~\eqref{dimensioniHubbleDec}) ed \`e all'origine del problema dell'orizzonte.

\begin{figure}
\centering
\begin{tikzpicture}

\fill[orange!5!white] (-4,.25) rectangle (4,3);

\fill[red!7!white] (-3,3)--(0,0)--(3,3)--cycle;
\draw[red] (-3,3)--(0,0)--(3,3);
\fill[red!7!white, path fading=fade up] (-3.5,3.5)--(-3,3)--(3,3)--(3.5,3.5)--cycle;
\draw[red, path fading=fade up] (-3.5,3.5)--(-3,3);
\draw[red, path fading=fade up] (3,3)--(3.5,3.5);

\fill[gray] (1.5,4) circle(2pt) node[black,anchor=south west]{$p$};
\fill[blue!5!white] (.5,3)--(1.5,4)--(2.5,3)--cycle;
\draw[blue] (.5,3)--(1.5,4)--(2.5,3);
\fill[blue!5!white, path fading=fade down] (.25,2.75)--(.5,3)--(2.5,3)--(2.75,2.75)--cycle;
\draw[blue, path fading=fade down] (.25,2.75)--(.5,3);
\draw[blue, path fading=fade down] (2.5,3)--(2.75,2.75);

%

\draw[help lines, black!80!white] (-4,3.25)--(4,3.25) node[above,near start, shift={(0,-3pt)}, gray]{\footnotesize\textsf{CMB}} node[black, anchor=west]{$t_{\textsf{dec}}$};
\draw[help lines] (-4,3)--(4,3) node[anchor=north west, near start, shift={(0,3pt)}]{\footnotesize\textsf{reheating}};

\draw[very thick, decorate, decoration={random steps,segment length=3pt,amplitude=1pt}] (-4,0)--(4,0) node[right]{$t_{\mathsf{i}}$};
\draw[help lines] (-4,.25)--(4,.25);
\draw[help lines] (-4,4)--(4,4) node[right, black]{$t_{0}$};

\draw[decorate,decoration=brace,shift={(-5pt,0)},gray,thick] (-4,0)--(-4,.25) node[midway, left, black]{\footnotesize\textsf{radiazione}};

\draw[decorate,decoration=brace,shift={(-5pt,0)},gray,thick] (-4,3.20)--(-4,4) node[midway, left, black]{\footnotesize\textsf{materia}};

\draw[decorate,decoration=brace,shift={(-5pt,3)},gray,thin] (-4,0)--(-4,.18) node[midway, left, black]{\footnotesize\textsf{radiazione}};


\fill (-4,1.625) node[black, anchor=west] {\footnotesize\textsf{INFLAZIONE}};
\draw[help lines,->] (-5.5,1.3)--(-4.7,1.3) node[below,black] {\footnotesize $\chi$};
\draw[help lines,->] (-5.5,1.3)--(-5.5,2.1) node[left,black] {\footnotesize $t$};
\fill[gray] (-5.5,1.3) circle(1.5pt);
%
\draw [help lines] (.5,3.1)--(.5,2.7);
\draw [help lines] (2.5,3.1)--(2.5,2.7);
\draw[help lines,<->] (0.5, 2.8) -- (2.5, 2.8) node[midway, below, black] {$\Delta\chi_{\textsf H}$};
\draw[help lines, dashed] (-3, 3.1)--(-3, -0.3);
\draw[help lines, dashed] (3, 3.1)--(3, -0.3);
\draw[help lines,<->] (-3, -0.2) -- (3, -0.2) node[midway, below, black] {$\Delta\chi_{\textsf{dec}}$};

\end{tikzpicture}
\caption{\textbf{Inflazione.} Una fase iniziale di espansione accelerata del fattore di scala avvicina $t_{\mathsf{dec}}$ a $t_0$ e porta il raggio comovente di Hubble ad essere inferiore alle dimensioni dell'orizzonte al momento del disaccoppiamento. Confrontare con l'analogo diagramma in assenza di inflazione in figura~\ref{fig:problemaorizzonte}.}
\label{fig:inflazione}
\end{figure}
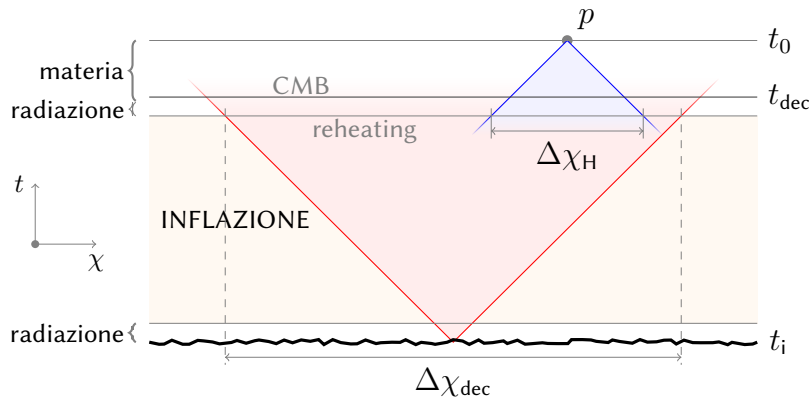

Questo comportamento si modifica ipotizzando che nei primi istanti della sua evoluzione l'universo abbia subito una fase di espansione accelerata, durante la quale, per l'equazione~\eqref{dchidtau}, il raggio comovente di Hubble \`e diminuito. Questo porta il momento di emissione della radiazione di fondo cosmico -- misurato in tempo conforme -- vicino al momento attuale, come illustrato dal diagramma conforme in figura~\ref{fig:inflazione}.
Infatti, si mostra che se la fase di espansione accelerata \`e sostenuta per un tempo sufficientemente lungo, si ottiene
\eq
\Delta\chi_{\textsf{dec}}=
\int^{\tau_{\mathsf{ric}}}_{\tau_{\mathsf{i}}}\!\frac{d\tau'}{a(\tau')}\gg\int_{\tau_{\mathsf{ric}}}^{\tau_0}\frac{d\tau'}{a(\tau')}=\Delta\chi_{\textsf{H}}.
\eeq
Questo risolve il problema dell'orizzonte, perch\`e la dimensione $\Delta\chi_{\textsf{dec}}$ (misurata in coordinate comoventi) delle regioni che hanno potuto mettersi in contatto causale e quindi termalizzare prima del disaccoppiamento sono molto grandi rispetto al raggio comovente di Hubble, e l'osservazione dell'isotropia della radiazione di fondo cosmica risulta dunque naturale: la radiazione che possiamo osservare proviene da punti dell'universo che hanno avuto tempo di mettersi in contatto causale prima di emetterla. Questo scenario ha anche il vantaggio di risolvere il problema della piattezza. Infatti, abbiamo visto nel paragrafo~\ref{sec:piattezza} che una fase di espansione accelerata sufficientemente lunga porta in modo naturale a delle condizioni iniziali con
$\Omega$ molto vicino ad uno. Una tale fase di espansione accelerata si dice fase di \textit{inflazione cosmica}\index{inflazione}.

Rimane da capire come generare questa fase di espansione accelerata dopo il Big Bang. Un fluido perfetto non \`e sufficiente: infatti abbiamo visto che sono le componenti del fluido perfetto con $w$ maggiore che determinano l'evoluzione primordiale del fattore di scala. Inizialmente l'evoluzione \`e dunque dominata dalla radiazione e l'espansione \`e decelerata, mentre l'espansione accelerata che misuriamo oggi, dovuta alla dominanza della costante cosmologica, \`e un fenomeno relativamente nuovo su scala cosmologica, e ininfluente per la discussione del problema dell'orizzonte.

Questa difficolt\`a pu\`o essere sormontata prendendo un campo di materia che non sia in regime idrodinamico, e dunque non riducibile a una componente di fluido perfetto. Questo campo, per conformarsi al principio cosmologico, deve essere omogeneo e non pu\`o individuare direzioni privilegiate, e dunque il suo tensore energia-impulso deve essere della forma $T^\mu{}_\nu=\mathrm{diag}\lp-\rho,P,P,P\rp$. In generale, non vi \`e un'equazione di stato che lega la densit\`a di energia $\rho$ alla pressione $P$. Tuttavia, vedremo che per un campo scalare, nel regime in cui possiamo trascurare l'energia cinetica del campo rispetto alla sua energia potenziale, vale approssimativamente la relazione $P\approx-\rho$. A tutti gli effetti tale materia \`e assimilabile all'energia di vuoto ($w=-1$) e la dinamica dell'universo si riduce effettivamente all'evoluzione guidata da una costante cosmologica. Questo meccanismo permette di innescare una fase di inflazione che, se sostenuta per un tempo sufficiente, risolve i problemi del modello cosmologico di Friedmann-Robertson-Walker. Il suo studio \`e l'oggetto del resto di questo capitolo.

In questa sezione poniamo $\hbar=1$ e $c=1$, e definiamo la massa di Planck ridotta\index{$\Mp$|see{massa di Planck ridotta}}\index{massa di Planck ridotta}
\eq
\Mp^2=\frac{1}{8\pi\newt}.
\eeq
In queste unit\`a, le lunghezze hanno dimensioni di una massa inversa, e l'azione \`e adimensionale.
Utilizzando i valori delle costanti fondamentali, troviamo che $\Mp=2.435\times10^{18}\,\mathsf{GeV/c^2}$.

\subsection{Un modello semplice di inflazione cosmica}

Consideriamo un campo scalare $\phi$ -- che chiameremo \textit{inflatone}\index{$\phi$|see{inflatone}}\index{inflatone} -- in un potenziale $V(\phi)$, minimalmente accoppiato alla gravit\`a. L'azione \`e la somma dell'azione di Einstein-Hilbert\index{azione di Einstein-Hilbert} e dell'azione di materia che descrive il campo scalare,
\begin{align}
S[g,\phi]&=\frac{\Mp^2}{2}\int d^4x\sqrt{-g}\,R
+ S_{\mathsf{M}}[g,\phi],\nonumber\\
S_{\mathsf{M}}[g,\phi]&=-\int d^4x\sqrt{-g}\lp\frac12\nabla_\mu\phi\nabla^\mu\phi+V(\phi)\rp.
\label{inflationAction}\end{align}
Il campo scalare ha perci\`o le dimensioni di una massa, $[\phi]=M$.

Richiedendo la stazionariet\`a di quest'azione rispetto a variazioni della metrica e del campo scalare,\footnote
{
Per effettuare le variazioni funzionali rispetto alla metrica, basta rammentare che la traccia della variazione del tensore di Ricci \`e una divergenza totale, e quindi non contribuisce alle equazioni del moto, e che per il determinante della metrica vale la relazione $\delta\lp\sqrt{-g}\rp = -\frac12 \sqrt{-g}\,g_{\mu\nu} \delta g^{\mu\nu}$ (vedi appendice~E.1 di~\cite{Wald:1984rg}).
}
otteniamo le equazioni di campo per il campo gravitazionale accoppiato con un campo scalare $\phi$,
\eq
G_{\mu\nu}=\frac1{\Mp^2}T_{\mu\nu},\qquad
\nabla^\mu\nabla_\mu\phi-V'(\phi)=0.
\eeq
La prima \`e l'equazione di Einstein, che ha per sorgente il tensore energia-impulso per il campo scalare\index{tensore energia-impulso!per un campo scalare} $\phi$
\eq
T_{\mu\nu}=-\frac2{\sqrt{-g}}\frac{\delta S_{\textsf{M}}}{\delta g^{\mu\nu}}=
\nabla_\mu\phi\nabla_\nu\phi-g_{\mu\nu}\left(\frac12(\nabla\phi)^2+V(\phi)\right),
\label{STphi}\eeq
mentre la seconda \`e l'equazione di campo per l'inflatone che generalizza l'equazione di Klein-Gordon\index{Klein-Gordon}. Infatti, se il campo scalare propaga su geometria curva, dobbiamo sostituire la metrica di Minkowski $\eta_{\mu\nu}$ e le derivate $\p_\mu$ con la metrica $g_{\mu\nu}$ dello spazio-tempo e le corrispondenti derivate covarianti $\nabla_\mu$, e laplaciano\index{laplaciano} (o d'Alembertiano) usuale $\eta^{\mu\nu}$ diventa l'operatore di Laplace associato a $g_{\mu\nu}$,
\eq
\Delta\phi\equiv\nabla^\mu\nabla_\mu\phi=\frac1{\sqrt{-g}}\p_\mu\lp\sqrt{-g}\,g^{\mu\nu}\p_\nu\phi\rp.
\eeq

Restringiamo queste equazioni a configurazioni spazialmente omogenee e isotrope, per rispettare il principio cosmologico. Inoltre, consideriamo solo sezioni spaziali piatte, con $k=+1$. La validit\`a di queste ipotesi durante l'epoca inflazionaria verr\`a discussa pi\'u avanti. La geometria deve dunque essere quella di Robertson-Walker. In questa sezione lavoriamo in coordinate cosmiche, e dunque la metrica assume la forma~\eqref{RWmetric}. Inoltre, per osservare il principio cosmologico, anche lo scalare $\phi$ deve essere spazialmente omogeneo, e dunque pu\`o dipendere solo dal tempo cosmico, $\phi=\phi(\tau)$. Sotto queste ipotesi, utilizzando $g^{\tau\tau}=-1$ e $\sqrt{-g}=a^3(\tau)\Sigma_k^2(\chi)\sin\theta$, troviamo
\eq
\Delta\phi=\frac1{a^3(\tau)}\p_\tau\lp a^3(\tau)\p_\tau\phi\rp=-\ddot\phi-3H\dot\phi,
\eeq
dove abbiamo introdotto il parametro di Hubble $H$ definito in~\eqref{c-Hubble_parm} e usiamo i punti per indicare derivate parziali rispetto al tempo cosmico, come di consueto. Allora l'equazione di campo per il campo scalare diventa
\eq
\ddot\phi+3H\dot\phi+V'(\phi)=0.
\label{eomphi}\eeq
Vediamo che il fattore di scala $a(\tau)$ entra nell'equazione dell'inflatone solo nel secondo termine, attraverso il parametro di Hubble che introduce uno smorzamento\index{termine di smorzamento} dipendente dal tempo.

Le componenti non nulle del tensore energia-impulso~\eqref{STphi} -- valutato per un campo scalare omogeneo sulla metrica di Robertson-Walker -- sono dunque
\beq
T^\tau{}_\tau=-\frac12\dot\phi^2-V(\phi),\qquad
T^\chi{}_\chi=T^\theta{}_\theta=T^\phi{}_\phi=\frac12\dot\phi^2-V(\phi),
\label{STscalar}\eeq
e formalmente si raccolgono nel tensore energia-impulso di un fluido perfetto~\eqref{Tfluidoperfetto}, con densit\`a di energia e pressione dipendenti da $(\phi,\dot\phi)$,
\eq
\rho=\frac12\dot\phi^2+V(\phi),\qquad
P=\frac12\dot\phi^2-V(\phi).
\label{scalarRhoP}\eeq
Questo non \`e sorprendente, e riflette semplicemente l'omogeneit\`a spaziale e l'isotropia che abbiamo imposto inizialmente. \`E invece importante sottolineare che non abbiamo un'equazione di stato, in quanto densit\`a di energia e pressione sono legate tra loro attraverso una dipendenza esplicita da $\phi$ e $\dot\phi$ che non si pu\`o eliminare: la dinamica del campo scalare non si riduce a quella di un fluido perfetto, nemmeno imponendo l'omogeneit\`a. Ci\`o nonostante, le equazioni di Einstein si riducono alle due equazioni di Friedmann~\eqref{F1} e \eqref{F2}, con $\rho$ e $P$ dati da~\eqref{scalarRhoP}.

Alternativamente, avremmo potuto ricavare l'equazione del moto per l'inflatone~\eqref{eomphi} dalla conservazione dell'energia~\eqref{econs} con $\rho$ e $P$ date da~\eqref{scalarRhoP}. L'origine del termine di smorzamento $3H\dot\phi$ \`e allora chiara: vi \`e un lavoro dovuto all'espansione dello spazio nel quale vive il fluido.

In conseguenza, solo due delle tre equazioni differenziali in gioco~\eqref{F1},~\eqref{F2}, e~\eqref{eomphi} sono indipendenti e -- dati i valori iniziali -- determinano completamente le due funzioni incognite $a(\tau)$ e $\phi(\tau)$. Scegliamo la prima equazione di Friedmann (con $k=+1$) e l'equazione per l'inflatone, ottenendo cos\'\i\ il sistema
\begin{align}
&\ddot\phi+3H\dot\phi=-V'(\phi),\\
&H^2=\frac1{3\Mp^2}\left[\frac12\dot\phi^2+V(\phi)\right].
\end{align}
Queste equazioni implicano la seconda equazione di Friedmann, che diventa
\eq
\frac{\ddot a}{a}=-\frac1{3\Mp^2}\lp\dot\phi^2-V(\phi)\rp,
\eeq
e mostra che per avere un'espansione accelerata dell'universo, $\ddot a>0$, \`e sufficiente che valga la relazione $V(\phi)>\dot\phi^2$. In particolare, se l'energia potenziale domina sulla sua energia cinetica,
\eq
V(\phi)\gg\frac12\dot\phi^2,
\label{sr1}\eeq
possiamo trascurare i termini dipendenti cinetici nel tensore energia-impulso \eqref{STscalar} e quest'ultimo si riduce a quello per l'energia di vuoto, con
\eq
\rho\approx V(\phi),\qquad P\approx-V(\phi).
\label{potDom}\eeq
In questo regime, il campo scalare si comporta come un fluido cosmologico con equazione di stato $P=-\rho$, ovvero $w=-1$. Siamo cos\'\i\ riusciti a generare una costante cosmologica effettiva che domina la dinamica cosmologica dell'universo primordiale. Finch\'e quest'approssimazione \`e valida, l'universo subisce un'espansione accelerata, con una crescita esponenziale del fattore di scala e una geometria di de~Sitter~\eqref{dSscale}. 
Per risolvere i problemi dell'orizzonte e della piattezza, questa fase inflativa deve tuttavia durare sufficientemente a lungo; abbiamo visto che abbiamo bisogno di una crescita di un fattore $10^{26}$ della scala dell'universo.

Per sostenere il regime inflazionario per un periodo sufficientemente lungo, mentre il campo scalare si muove verso il minimo del potenziale $V(\Phi)$, la sua energia cinetica deve variare poco, per mantenere~\eqref{potDom}. In altre parole, richiediamo che l'accelerazione $\ddot\phi$ sia trascurabile rispetto al termine di smorzamento,
\eq
|\ddot\phi|\ll3H|\dot\phi|.
\label{sr2}\eeq
in modo tale che l'equazione di campo per l'inflatone diventi approssimativamente
\eq
3H\dot\phi\approx-V'(\phi).
\label{eqPhiApprox}\eeq
Un tale moto si dice \textit{sovrasmorzato}\index{moto sovrasmorzato}\index{sovrasmorzato|see{moto sovrasmorzato}}, analogo al moto di un corpo in un fluido viscoso. Il campo $\phi$ parte da un valore elevato del potenziale e rotola lentamente verso il minimo, generando effettivamente una costante cosmologica che guida l'espansione accelerata. Questo regime di moto sovrasmorzato, guidato dall'energia potenziale, si dice di \textit{slow-roll}\index{slow-roll}.
Quando le \emph{condizioni di slow-roll}\index{slow-roll!condizioni di} sono soddisfatte, le equazioni del moto si riducono al sistema
\bseq\begin{align}
3H\dot\phi&=-V'(\phi),
\label{sreq1}\\
H^2&=\frac1{3\Mp^2}V(\phi).
\label{sreq2}
\end{align}\eseq
Verifichiamo che le condizioni di \emph{slow-roll} sono compatibili fra loro, in quanto dalla seconda equazione segue che se il valore del potenziale \`e grande, anche il parametro di Hubble lo \`e, e la prima equazione ci assicura dunque che $\dot\phi$ \`e piccolo, per lo meno se il potenziale $V(\phi)$ non \`e troppo ripido.

\`E possibile riformulare le condizioni di \emph{slow-roll} come condizioni sulla forma del potenziale $V(\phi)$. Infatti, sotituendo la seconda equazione~\eqref{sreq2} nel quadrato della prima~\eqref{sreq1}, troviamo che
\eq
\dot\phi^2=\frac{V'^2}{9H^2}=\frac{\Mp^2}3\frac{V'^2}{V}
\eeq
e la condizione~\eqref{sr1} diventa semplicemente $(V'/V)^2\ll1$, ovvero che il potenziale non sia troppo ripido.
\definizione{Il primo parametro di slow-roll\index{slow-roll!primo parametro di} $\ep$\index{$\ep$|see{slow-roll, primo parametro di}} \`e dato da
\eq
\ep=\frac{M_{\mathrm{p}}^2}2\lp\frac{V'}V\rp^2,
\eeq
e l'universo \`e in una fase inflazionaria, guidata dall'energia potenziale dell'inflatone, quando \`e verificata la prima condizione di slow-roll $\ep\ll1$.
}
Per esprimere la condizione di sovrasmorzamento in termini del potenziale, scriviamo l'equazione~\eqref{sreq1} per l'inflatone come $\dot\phi=-V'(\phi)/3H$, e ne prendiamo la derivata temporale. La relazione risulatante si semplifica utilizzando l'equazione di Friedmann~\eqref{sreq2} e la sua derivata temporale, ottenendo cos\`\i
\eq
\ddot\phi=-\lp\ep+\Mp^2\frac{V''}{V}\rp H\dot\phi.
\eeq
Per verificare la seconda condizione di \emph{slow-roll}~\eqref{sr2}, il termine fra parentesi deve essere molto piccolo. Se la prima condizione di \emph{slow-roll} $\ep\ll1$ \`e verificata, \`e sufficiente imporre la condizione supplementare $\Mp^2V''/V\ll1$.
\definizione{
Il secondo parametro di \emph{slow-roll}\index{slow-roll!secondo parametro di} $\eta$\index{$\eta$|see{slow-roll, secondo parametro di}} \`e dato da 
\beq
\eta=\Mp^2\frac{V''}{V}\,,
\eeq
e l'evoluzione dell'inflatone \`e guidata dal termine di attrito se $|\eta|\ll1$.
}
Abbiamo cos\'\i\ ridotto le condizioni di \emph{slow-roll}\index{slow-roll!condizioni di} a condizioni sulla forma del potenziale. Se per il valore iniziale del campo $\ep\ll1$, la dinamica cosmologica \`e dominata dal potenziale dell'inflatone e si innesca l'inflazione. Se inoltre $|\eta|\ll1$, l'evoluzione successiva \`e sovrasmorzata e questa fase inflazionaria viene sostenuta per tempo lungo.

\begin{Esempio}
\textbf{Potenziali polinomiali:} Per un potenziale polinomiale $V\propto\phi^n$, abbiamo $V'\propto V/\phi$ e $V''\propto V/\phi^2$, e dunque i parametri di \emph{slow-roll} sono dati da $\ep\propto\Mp^2/\phi^2$ e $\eta\propto\Mp^2/\phi^2$. Le due condizioni $\ep\ll1$ e $|\eta|\ll1$ sono dunque sempre soddisfatte se $\phi$ \`e sufficientemente grande rispetto alla scala di Planck $\Mp$.
\end{Esempio}

\paragraph{Osservazione:} Mentre le condizioni di \emph{slow-roll} $\ep\ll1$ e $|\eta|\ll1$ sono necessarie, affinch\'e avvenga l'inflazione, non sono sufficienti per scatenarla.
Tuttavia, \`e sufficiente verificare che inizialmente $\dot\phi$ verifica l'equazione~\eqref{eqPhiApprox}.

\subsection{La durata dell'inflazione}
Per risolvere i problemi dell'orizzonte e della piattezza, l'inflazione deve durare sufficientemente a lungo, e $a(\tau)$ deve crescere di un fattore $10^{26}$ prima di terminare e iniziare la fase di \textit{Big Bang caldo}\index{Big Bang!caldo} (o \emph{hot Big Bang}). Per misurare la quantit\`a di inflazione avvenuta, \`e comodo definire il \textit{numero di $e$-foldings}. 
\definizione{Il numero di $e$-foldings~$N$
\index{$N$|see{$e$-foldings, numero di}}
\index{$e$-foldings!numero di}
\index{numero di $e$-foldings|see{$e$-foldings}}
\`e il logaritmo del fattore di cui si \`e espanso l'universo dall'inizio dell'inflazione $\tau_i$ alla fine della fase inflazionaria $\tau_f$,
\beq
N=\ln\frac{a(\tau_f)}{a(\tau_i)}\,.
\eeq
}
Anche questa quantit\`a dipende solamente dalla forma del potenziale $V(\phi)$ e dal valore iniziale $\phi_i$ e finale $\phi_f$ dell'inflatone, ma non dai dettagli della sua evoluzione. Infatti, combinando le equazioni in regime di \emph{slow-roll}~\eqref{sreq1} e \eqref{sreq2}, otteniamo
\eq
H=-\frac1{\Mp^2}\frac{V(\phi)}{V'(\phi)}\dot\phi,
\eeq
ed esprimendo $N$ come integrale temporale troviamo
\eq
N=\int_{\tau_i}^{\tau_f}\frac{\dot a}{a}\,d\tau
=\int_{\tau_i}^{\tau_f}H\,d\tau
=-\frac1{\Mp^2}\int_{\tau_i}^{\tau_f}\frac{V(\phi)}{V'(\phi)}\dot\phi\,d\tau.
\label{H}\eeq
Cambiando variabile di integrazione (supponendo che $\phi(\tau)$ sia una funzione monotona del suo argomento) otteniamo finalmente
\beq
N=-\frac{1}{\Mp^2}\int^{\phi_f}_{\phi_i}\frac{V(\phi)}{V'(\phi)}\,d\phi\,.
\label{N}
\eeq
Abbiamo visto che per risolvere i problemi dell'orizzonte e della piattezza, durante l'era inflazionaria le dimensioni dell'universo devono crescere di un fattore $10^{26}$. Prendendo il logaritmo naturale di questo numero, troviamo che la quantit\`a minima di inflazione di cui abbiamo bisogno corrisponde a un numero di $e$-foldings $N\approx60$.

\begin{Esempio}
\textbf{Inflazione con potenziale polinomiale:} Consideriamo il potenziale seguente
\eq
V(\phi)=\mu^{4-p}\phi^p,
\label{potPoly}\eeq
dove $\mu$ \`e una costante con dimensioni di una massa, e $p>0$ \`e l'esponente. L'inflazione termina quando l'inflatone raggiunge il minimo del potenziale, $\phi\approx0$. Dall'eq.~\eqref{N}, il numero di $e$-foldings \`e dato dal valore iniziale $\phi_i$ dell'inflatone,
\eq
N=-\frac1{\Mp^2}\int^0_{\phi_i}\frac\phi{p}d\phi=\frac1{2\Mp^2}\frac{\phi_i^2}{p}.
\eeq
Vediamo dunque che per avere $N\approx60$ $e$-foldings, il valore iniziale dell'inflatone deve essere
\eq
|\phi_i|>\Mp\sqrt{2N n}\approx15\Mp.
\eeq
\end{Esempio}
Durante la fase inflazionaria in regime di \emph{slow-roll}, l'evoluzione del fattore di scala \`e approssimato da un'esponenziale. Tuttavia, mentre il dilatone si avvicina al minimo del potenziale, il valore del potenziale -- e dunque della costante cosmologica effettiva $\Lambda_{\textsf{eff}}$ -- diminuisce. Vi sono dunque deviazioni dalla geometria di de~Sitter, che comunque approssima bene la geometria dell'universo su periodi in cui $\Lambda_{\textsf{eff}}$ varia poco.
\begin{Esercizio} \textbf{Potenziale quadratico} \quad
Si consideri un potenziale quadratico, della forma~\eqref{potPoly} con $p=2$. Risolvere le equazioni di evoluzione~\eqref{sreq1} e~\eqref{sreq2} in regime di \emph{slow-roll}, e trovare la dipendenza dal del fattore di scala $a(\tau)$ dal tempo. Su che scale di tempi si verifica una deviazione dalla crescita esponenziale di de~Sitter? Qual'\`e la durata dell'inflazione se l'inflatone ha la massa del bosone di Higgs\index{Higgs!bosone di}?
\end{Esercizio}
I dati osservativi provenienti dalla radiazione cosmica di fondo sono in ottimo accordo con le previsioni dell'inflazione in regime di \emph{slow-roll} con un singolo campo scalare $\phi$, come quello che abbiamo descritto in questa sezione. Queste osservazioni sono anche in grado di discernere tra diversi potenziali $V(\phi)$, e attualmente favoriscono -- rispetto ai potenziali polinomiali -- i potenziali di tipo \textit{plateau},
\index{plateau}
ovvero con una parte molto piatta che permette di sostenere l'inflazione per tempi lunghi e con costante cosmologica effettiva praticamente costante. Le osservazioni dei prossimi anni (e in particolare le osservazioni della polarizzazione della radiazione cosmica di fondo) ci permetteranno di misurare dettagli del potenziale dell'inflatone. Un modello popolare con potenziale di plateau \`e il modello di Starobinsky, che vediamo nell'esercizio seguente.
\begin{Esercizio}
\textbf{Il modello di Starobinsky}
\index{Starobinsky, modello di}
\quad
Aggiungendo una correzione quadratica nello scalare di Ricci $R$ della forma $\alpha R^2/2\Mp^2$ nell'azione di Einstein-Hilbert, ed effettuando una trasformazione conforme, si ottiene una teoria di gravit\`a accoppiata a un campo scalare $\phi$ descritta dall'azione~\eqref{inflationAction}, con potenziale di tipo \emph{plateau} dato da
\eq
V(\phi)=\frac{\Mp^4}{4\alpha}\lp1-\exp\lp-\sqrt{\frac23}\frac\phi\Mp\rp\rp^2.
\eeq
Per questo motivo si parla di \emph{inflazione $R^2$ di Starobinsky}.
Trovare i parametri di \emph{slow-roll} $\ep$ e $\eta$. Per che valori di $\phi$ si svolge l'inflazione? E quale \`e il valore minimo della costante $\alpha$ affinch\'e ci\`o accada? Trovare il numero di $e$-foldings come funzione del valore iniziale $\phi_i$ dell'inflatone, e la durata dell'inflazione.
\end{Esercizio}

\subsection{Reheating e Big Bang caldo}
\index{reheating}
\index{Big Bang!caldo}
Durante la fase di inflazione, tutta la materia -- ad eccezione del campo scalare $\phi$ -- subisce uno spostamento verso il rosso molto elevato. I fluidi cosmologici vengono diluiti a densit\`a bassissime, e la loro temperatura precipita, portando a una soppressione delle disomogeneit\`a iniziali.
Questa fase continua fintanto che i parametri di slow-roll rimangono piccoli. Quando $\epsilon\approx\eta\approx1$, l'evoluzione dell'inflatone esce dal regime sovrasmorzato, e passa a un moto oscillatorio smorzato intorno al minimo del potenziale, durante il quale il campo inflatonico dissipa la sua energia residua decadendo nelle particelle del modello standard e riscaldando l'universo. Questo processo genera le condizioni iniziali per le densit\`a di particelle e temperatura per l'evoluzione successiva dell'universo. Questo \`e il Big Bang caldo, dal quale scaturisce l'evoluzione standard cosmologica che abbiamo studiato durante le prime lezioni, con la fase iniziale dominata dalla radiazione, seguita dalla fase dominata dalla materia.

\subsection{L'attrattore inflazionario}

L'universo, attraverso il meccanismo dell'inflazione, raggiunge naturalmente uno stato dalle propriet\`a tali che l'evoluzione successiva (universo dominato da radiazione prima, dalla materia poi), porta all'universo che osserviamo oggi, con propriet\`a compatibili alle osservazioni cosmologiche.

Ma abbiamo davvero risolto il problema di aggiustamento
\index{aggiustamento fine}
fine che avevamo individuato nello studio dell'orizzonte di Hubble e nell'osservazione della piattezza spaziale? In fondo non abbiamo fatto altro che sostituire le condizioni iniziali del Big Bang caldo con condizioni iniziali prima dell'innescarsi dell'inflazione, e il rischio \`e di aver semplicemente nascosto l'aggiustamento fine delle condizioni iniziali dietro alla fase inflazionaria.

Per rispondere a questa domanda dobbiamo capire quanto fortemente lo stato dell'universo all'uscita dell'inflazione dipenda dalle condizioni iniziali che imponiamo, e questo lo facciamo studiando la stabilit\`a delle soluzioni delle equazioni che determinano la dinamica dell'inflazione.

Supponendo che $\phi(\tau)$ sia una funzione monotona del tempo $\tau$,
possiamo utilizzare il valore assunto dall'inflatone come variabile temporale, e riformulare le equazioni di Friedmann~\eqref{F1}-\eqref{F2} come un sistema di equazioni per il parametro di Hubble $H(\phi)$,
\begin{subequations}
\begin{align}
&H'(\phi)^2-\frac3{2\Mp^2}H^2(\phi)=-\frac1{2\Mp^4}V(\phi),
\label{FH1}
\\
&\dot\phi=-2\Mp^2H'(\phi),
\label{FH2}
\end{align}
\end{subequations}
dove con l'apostrofo indichiamo la derivata rispetto a $\phi$.
Risolvendo la prima equazione, ora disaccoppiata, determiniamo la soluzione $H(\phi)$ che identifica una soluzione cosmologica. Infatti, da una tale soluzione, possiamo usare la seconda equazione~\eqref{FH2} per trovare la dipendenza $\phi(\tau)$ dell'inflatone dal tempo, e dunque anche l'evoluzione $a(\tau)$ del fattore di scala.

Supponiamo ora di avere una soluzione $\bar H(\phi)$ dell'equazione~\eqref{FH1} e consideriamo una sua piccola perturbazione,
\eq
H(\phi) = \bar H(\phi) + \delta H(\phi).
\eeq
Sostituendo nell'equazione~\eqref{FH1} e linearizzando rispetto alla perturbazione, troviamo un'equazione differenziale separabile per $\delta H$,
\eq
\frac{\delta H'}{\delta H}=\frac{3}{\Mp}\frac{\bar H'}{\bar H},
\eeq
la cui soluzione \`e data da
\eq
\delta H(\phi)=\delta H(\phi_i)\exp\left(
\frac{3}{\Mp^2}\int_{\phi_i}^\phi\frac{\bar H}{\bar H'}d\phi
\right).
\eeq
Dalla relazione~\eqref{FH2} sappiamo che $\dot\phi$ e $H'(\phi)$ hanno segni opposti, e dunque l'argomento dell'esponenziale nella formula qui sopra \`e negativo: la pertubazione iniziale viene smorzata molto rapidamente dalla dinamica inflazionaria.
Possiamo legare l'integrale al membro di destra al numero di $e$-foldings utilizzand~\eqref{H},
\eq
N=\int_{\tau_i}^{\tau_f}\bar H\,d\tau=\int_{\phi_i}^{\phi_f}\frac{\bar H}{\dot\phi}\,d\phi,
\eeq
che, ricordando la relazione~\eqref{FH2}, diventa
\eq
N=-\frac1{2\Mp^2}\int_{\phi_i}^\phi\frac{\bar H}{\bar H'}.
\eeq
Una perturbazione iniziale $\delta H(\phi_i)$, al termine dell'inflazione \`e dunque attenuata esponenzialmente nel numero di $e$-foldings,
\eq
\delta H(\phi)=\delta H(\phi_i)\,e^{-6N}.
\eeq
Prendendo $N=60$ per la durata dell'inflazione, vediamo che una qualunque perturbazione viene soppressa per un fattore $10^{-360}$, ovvero non ne rimane pi\'u traccia dopo il reheating!
Le soluzioni inflazionarie sono dunque attrattori\index{attrattore inflazionario}: e tutte le soluzioni si avvicinano molto rapidamente l'una all'altra fino al punto di risultare indistinguibili fra loro all'uscita dell'inflazione. \`E importante osservare che non vi \`e un singolo attrattore. Il ragionamento precedente \`e indipendente dalla scelta della soluzione $\bar H$ che si perturba, e dunque tutte le soluzioni sono attrattori per tutte le altre soluzioni. In conclusione, se il potenziale $V(\phi)$ ammette una soluzione inflazionaria, allora l'inflazione \`e inevitabile, e il valore dei campi all'uscita dell'inflazione sono insensibili ai dettagli delle condizioni iniziali per l'inflazione. Il problema dell'aggiustamento fine \`e cos\'\i\ risolto.


\appendix

\chapter{Esercizi supplementari}

\begin{Esercizio}\label{es:redshift}
\textbf{Il redshift cosmologico} \textsf{(Wald~\cite{Wald:1984rg})}\quad
Si consideri un modello cosmologico di Robertson-Walker con metrica
\eq
ds^2=-d\tau^2+a^2(\tau)\lp d\chi^2+\Si_\kappa^2(\chi)d\Om^2\rp,
\eeq
dove $\kappa=+1$, $0$, o $-1$ per un universo chiuso, piatto o aperto, rispettivamente.
Vogliamo mostrare la formula del redshift cosmologico, indipendentemente dalla curvatura $\kappa$ delle sezioni spaziali.
Chiamiamo $u^\mu$ il vettore quadrivelocit\`a degli osservatori isotropi $u=\p_\tau$, definiamo il proiettore $h$ sulle sezioni a $\tau$ costante $h_{\mu\nu}=g_{\mu\nu}+u_\mu u_\nu$ e indichiamo con un punto la derivata rispetto a $\tau$, $\dot a=da/d\tau$.
\begin{questions}
\item Mostrare che $\nabla_\mu u_\nu=(\dot a/a)h_{\mu\nu}$.
\item Mostrare che lungo le geodetiche nulle $x^\mu=x^\mu(\la)$ la frequenza $\omega$ del segnale, misurata dagli osservatori isotropi, varia secondo,
\eq
\frac{d\omega}{d\lambda}=-k^\mu k^\nu\nabla_\mu u_\nu=-\frac{\dot a}{a}\om^2
\label{domega}\eeq
con $\la$ parametro affine della geodetica, $k^\mu$ 
il suo vettore tangente e $\omega=-k^\mu u_\mu$.
\item Utilizzando \eqref{domega}, dedurre la formula per il redshift cosmologico,
\eq
\frac{\om_2}{\om_1}=\frac{a(\tau_1)}{a(\tau_2)}.
\label{redshiftcosmo}\eeq
\end{questions}
\end{Esercizio}

\begin{Esercizio}
\textbf{Inflazione caotica con potenziale quadratico}
Si consideri un modello di inflazione con un singolo inflatone $\phi$ accoppiato minimalmente al campo gravitazionale, e con potenziale quadratico $V(\phi)=\mu^2\phi^2$. Misuriamo i tempi con il tempo proprio $\tau$ degli osservatori isotropi e chiamiamo $\phi_{\mathsf{i}}$ il valore dell'inflatone al tempo iniziale $\tau_{\mathsf{i}}$ dell'inflazione, e $\phi_{\mathsf{f}}$ il suo valore finale al termine $\tau_{\textsf{f}}$ dell'epoca inflazionaria. Indichiamo la durata dell'inflazione con $\Delta\tau=\tau_{\textsf{f}}-\tau_{\mathsf{i}}$.
\begin{questions}
\item Trovare i parametri di \emph{slow-roll} $\ep$ e $\eta$ per questo potenziale. Quale \`e la condizione che deve soddisfare l'inflatone $\phi$ affinch\'e valga l'approssimazione di \emph{slow-roll}?
\item Per quale valore $\phi_f$ l'evoluzione esce dal regime di \emph{slow-roll} e inizia la fase di \emph{reheating}?
\item Esprimere il numero di $e$-foldings $N=\ln(a_{\textsf{f}}/a_{\textsf{i}})$ come funzione di $\phi_{\textsf{i}}$ e $\phi_{\textsf{f}}$. 
Come dipende il valore iniziale $\phi_{\mathsf{i}}$ dell'inflatone da $N$? Quanto vale, in unit\`a di $\Mp$, per $N\approx60$?
\item Risolvere il sistema di equazioni per l'evoluzione del fattore di scala $a(\tau)$ e dell'inflatone $\phi(\tau)$ in regime di \emph{slow-roll}, e trovare la dipendenza esplicita del fattore di scala $a(\tau)$ dal tempo cosmico $\tau$.
\item Come dipende la durata $\Delta\tau$ dell'inflazione da $N$ e $\mu$? Calcolare $\Delta\tau$ in secondi, se l'inflatone ha la massa del bosone di Higgs.
\item \`E corretto assimilare l'inflazione con una fase di espansione esponenziale di tipo de~Sitter? Su che scale di tempi si verifica una deviazione dall'evoluzione quasi-de~Sitter?
\end{questions}
\noindent\emph{Nota: Possono risultare utili i seguenti valori della massa di Planck ridotta $\Mp$, della lunghezza di Planck $\ell_{\mathsf{p}}$, del tempo di Planck $t_{\mathsf{p}}$, e della massa $M_{\mathsf{Higgs}}$ del bosone di Higgs,
\begin{align*}
\Mp&=\sqrt{\frac{\hbar c}{8\pi\mc G}}\approx2.435\times10^{18}\,\mathsf{GeV}/\mathsf{c}^2,\qquad&
M_{\mathsf{Higgs}}&=125\,\mathsf{GeV}/\mathsf{c}^2
\\
\ell_{\mathsf{p}}&=\sqrt{\frac{\hbar\mc G}{c^3}}\approx 1.6\times10^{-33}\,\mathsf{cm}\qquad&
t_{\mathsf{p}}&=\sqrt{\frac{\hbar\mc G}{c^5}}\approx 5.4\times10^{-44}\,\mathsf{s}
\end{align*}
}
\end{Esercizio}

\newenvironment{acknowledgements}%
    {\cleardoublepage\thispagestyle{empty}\null\vfill\begin{center}%
    \bfseries Ringraziamenti\end{center}}%
    {\vfill\null}
        \begin{acknowledgements}
        Questi appunti sono frutto di lezioni tenute dal 2010 presso l'Universit\`a
        degli Studi di Milano, nell'ambito della Laurea Magistrale in Fisica.

        \`E un grande piacere ringraziare Dietmar Klemm e Alberto Santambrogio
        per l'ospitalit\`a, sempre eccellente!
        Ringrazio anche tutti gli studenti che, con le loro domande
        e commenti, hanno contribuito a migliorare questi appunti.
        
        Desidero inoltre ringraziare il Dipartimento di Fisica dell'Universit\`a di Milano
        e la sezione di Milano dell'INFN per l'ospitalit\`a e per il supporto
        finanziario.
        Questo lavoro \`e stato parzialmente supportato dalla
        Marie Curie Intra-European Fellowship nr 628104 nell'ambito del
        7th European Community Framework Programme FP7/2007-2013.

        \end{acknowledgements}


\printindex

\end{document}